\documentclass[conference]{IEEEtran}
\IEEEoverridecommandlockouts
\usepackage{cite}
\usepackage{amsmath,amssymb,amsfonts}
\usepackage{graphicx}
\usepackage{textcomp}
\usepackage{xcolor}
\usepackage[dvipsnames]{xcolor}

\usepackage{float}
\makeatletter

\newcommand\floatc@ruledBlack[2]{{\color{black}\bfseries #1:} {\color{black}#2}}

\renewcommand\fs@ruled{%
  \def\@fs@cfont{\bfseries}\let\@fs@capt\floatc@ruledBlack
  
  \def\@fs@pre{{\color{black}\hrule height.8pt depth0pt \kern2pt}}%
  \def\@fs@post{{\color{black}\kern2pt\hrule\relax}}%
  \def\@fs@mid{{\color{black}\kern2pt\hrule\kern2pt}}%
  
  \let\@fs@iftopcapt\iftrue
  \def\@fs@beg{\color{black}}
}

\makeatother
\definecolor{R1}{named}{blue}          % Reviewer #1
\definecolor{R2}{named}{ForestGreen}   % Reviewer #2
\definecolor{R4}{named}{BrickRed}    % Reviewer #4
\definecolor{R5}{named}{Purple}       % Reviewer #5

\usepackage{algpseudocode}
\usepackage{algorithm}
\usepackage{amsmath}
\usepackage{authblk} % For author affiliation
\usepackage{tikz}
\usepackage{subfigure}
\usepackage{url}
\usepackage{balance}
\usepackage{flushend}
\usepackage[hidelinks]{hyperref}

\usepackage[normalem]{ulem}

\def\BibTeX{{\rm B\kern-.05em{\sc i\kern-.025em b}\kern-.08em
    T\kern-.1667em\lower.7ex\hbox{E}\kern-.125emX}}
\begin{document}

\title{Nezha: A Key-Value Separated Distributed Store with Optimized Raft Integration\\
% {\footnotesize \textsuperscript{*}Note: Sub-titles are not captured for https://ieeexplore.ieee.org  and
}

\author[1,2]{Yangyang Wang}
\author[1,2]{Yucong Dong}
\author[1,2]{Ziqian Cheng}
% \author[2]{Yunpeng Chai}
\author[1,2*]{Zichen Xu}
\affil[1]{School of Artificial Intelligence, Nanchang University, China}
\affil[2]{School of Mathematics and Computer Sciences, Nanchang University, China}
\affil[$^*$]{Corresponding Author: xuz@ncu.edu.cn}

% \and
% \and
% \and
% \and
% \and

\maketitle

\begin{abstract}

Distributed key-value stores are widely adopted to support elastic big data applications, leveraging purpose-built consensus algorithms like Raft to ensure data consistency. However, through systematic analysis, we reveal a critical performance issue in such consistent stores, i.e., overlapping persistence operations between consensus protocols and underlying storage engines result in significant I/O overhead. To address this issue, we present \textit{Nezha}, a prototype distributed storage system that innovatively integrates key-value separation with Raft to provide scalable throughput in a strong consistency guarantee. \textit{Nezha} redesigns the persistence strategy at the operation level and incorporates leveled garbage collection, significantly improving read and write performance while preserving Raft's safety properties.
Experimental results demonstrate that, on average, \textit{Nezha} achieves throughput improvements of 460.2\%, 12.5\%, and 72.6\% for put, get, and scan operations, respectively.
\end{abstract}


\section{Introduction}
With the explosive growth of big data applications, distributed key-value storage systems have become the indispensable infrastructure for modern data-intensive applications \cite{b19,b20,b21,b22}. Key-value stores offer a simple yet powerful approach to data management, directly supporting large-scale applications across diverse domains, including e-commerce \cite{b23}, social networks \cite{b24}, and machine learning \cite{b25}. Among various implementations, systems based on log-structured merge-trees (LSM-tree) \cite{b26} have gained widespread adoption due to their ability to optimize write operations through sequential disk I/O \cite{b20,b21,b22,b45,b46}. However, this write-optimized design inevitably comes at the cost of read performance: the multi-level structure of LSM-tree requires traversing multiple files even for simple point queries \cite{b12}. Meanwhile, the write amplification introduced by the compaction process significantly reduces storage efficiency \cite{b12,b47}.
\par
Distributed storage systems face the dual challenges of maintaining both high performance and data consistency. The latter challenge is commonly addressed through distributed consensus protocols, particularly Paxos \cite{b1,b2} and Raft \cite{b3,b4}, which ensure consistent data replication across multiple nodes. Raft has emerged as the preferred protocol in production systems, including etcd \cite{b27}, CockroachDB \cite{b36}, TiDB \cite{b37}, and PolarFS \cite{b38}, due to its understandable design and straightforward implementation.
However, Raft's safety guarantees require persistent storage of client requests, resulting in redundant disk writes when implemented in distributed key-value storage systems.
\par
\textbf{\textit{Summary of Findings.}} In Raft-based distributed key-value stores using LSM-tree \cite{b26} storage engines, each data write operation requires at least three distinct disk writes: one for Raft log persistence, another for LSM-tree's Write-Ahead-Log (WAL), and a third for Memtable persistence. Additional disk write overhead comes from the LSM-tree's background compaction process, which continuously merges and rewrites data across multiple levels. While these redundant write operations ensure data durability, our analysis reveals that such comprehensive persistence is not always necessary for all data, suggesting potential opportunities for performance optimization.
\par
\textbf{\textit{Overview of This Work}}. To address these redundant write operations while preserving data consistency guarantees, we present \textit{Nezha}, a novel distributed key-value storage system that achieves comprehensive performance improvements through architectural innovations. Our work encompasses the following innovation designs: 1) For write requests, we propose KVS-Raft, a consensus algorithm specifically optimized for distributed key-value storage by organically \textit{integrating a key-value separation strategy with the Raft protocol}. This integration reduces write amplification while maintaining Raft's safety properties; 2) For read requests, we design a \textit{Raft-aware Garbage Collection (GC) framework} to balance read performance; 3) For request processing correctness, we implement a \textit{three-phase request processing mechanism} that ensures correct handling of read and write requests across different optimization phases of the system.
\par
To validate our design, we implemented Nezha and conducted comprehensive experimental evaluations. Experimental results show substantial performance improvements over a traditional Raft-based key-value storage system: \textit{the average throughput increases by 460.2\%, 12.5\%, and 72.6\% for put, get, and scan operations, respectively.}
\par
The main contributions of this paper include:
\begin{itemize}%[topsep=1pt]
    \item We identify that traditional consensus protocol implementations in distributed key-value storage systems introduce unnecessary data persistence operations, which have become performance bottlenecks in modern hardware environments.
    \item We design Nezha, a novel distributed key-value storage system. At the consensus layer, we propose KVS-Raft, an enhanced consensus protocol that integrates key-value separation while maintaining strong consistency. At the storage layer, we design a Raft-aware GC framework complemented by a three-phase request processing mechanism to ensure correct request handling during system optimization.
    \item We implement Nezha and conduct extensive experimental evaluations, demonstrating significant performance improvements over traditional Raft-based storage systems across various workload patterns.
\end{itemize}
The remainder of this paper is organized as follows. Section \ref{sectio2} introduces the background and motivations. Section \ref{sectio3} presents the design of Nezha. Section \ref{sectio4} describes the implementation of the system and the evaluations. Section \ref{sectio5} discusses related work. Finally, Section \ref{sectio6} concludes the paper.

\section{BACKGROUND AND MOTIVATION}
\label{sectio2}
% This section establishes the foundation for our research. We begin with basic concepts of distributed consensus and data persistence in Section \ref{section2.1}, followed by a comprehensive review of the Raft protocol in Section \ref{section2.2}. In Section \ref{section2.3}, we identify key performance bottlenecks in current distributed key-value storage systems, particularly the redundant data persistence problem, and explore the potential benefits of combining key-value separation with Raft.
This section establishes the foundation for our research. We begin with basic concepts of distributed consensus and data persistence in Section \ref{section2.1}, followed by a comprehensive review of the Raft protocol in Section \ref{section2.2}. 
Section \ref{section2.3} introduces key-value separation, a fundamental technique that inspires our design. In Section \ref{section2.4}, we identify key performance bottlenecks in current distributed key-value storage systems, particularly the redundant data persistence problem, and explore the potential benefits of combining key-value separation with Raft.

\subsection{Consensus Algorithms and Data Persistence}
\label{section2.1}
\textbf{Definition 2.1}. (Distributed Consensus) \textit{Distributed consensus is a fundamental problem in distributed systems that requires multiple nodes in a system to reach agreement on a value or state, even in the presence of node failures, network partitions, and other fault scenarios.}
\par
\textbf{Definition 2.2}. (Data Persistence) \textit{Data persistence refers to the process of transferring data from volatile storage (e.g., memory) to non-volatile storage devices (e.g., SSDs, HDDs). This process relies on file system write operations and synchronization mechanisms to ensure data durability and consistency semantics.}
\par

Distributed consensus serves as the cornerstone for building reliable distributed systems. Across various distributed system applications, ensuring all nodes reach agreement on critical data and operations is paramount \cite{b29,b30,b31}. For instance, distributed storage systems must ensure nodes agree on the total order of data modification operations \cite{b32}. This consistency requirement directly impacts system correctness and availability.
\par
To maintain such consistency guarantees, data persistence is a crucial mechanism for ensuring reliability. While modern storage systems typically employ memory caching to enhance performance \cite{b33,b34,b35}, the volatile nature of memory necessitates periodic persistence operations to prevent data loss due to power failures or system crashes. This process usually involves system calls such as fsync(), which introduce significant I/O overhead. Therefore, balancing the frequency of persistence operations against performance overhead becomes a critical consideration in system design.
\par
In practical system implementations, these two fundamental requirements - distributed consensus and data persistence - are tightly coupled. To ensure system consistency after failure recovery, consensus algorithms typically require persistent storage of agreed-upon data. This persistence mechanism not only supports state recovery after node restarts but also provides the foundation for data synchronization when new nodes join the system. While this ``consensus-persistence" coupling pattern has become a standard paradigm in distributed system design, it also presents significant performance challenges.

\subsection{Raft protocol}
\label{section2.2}
Raft \cite{b3} significantly simplifies the complexity of distributed consensus by decomposing it into three relatively independent subproblems: Leader Election, Log Replication, and Safety Guarantee. This design approach has led to its widespread adoption in many practical systems, such as etcd \cite{b27}, CockroachDB \cite{b36}, TiDB \cite{b37}, and PolarFS \cite{b38}.
% Among various consensus protocols designed to address these challenges, Raft \cite{b3} has emerged as a particularly influential solution due to its practical design philosophy. Raft significantly simplifies the complexity of distributed consensus by decomposing it into three relatively independent subproblems: Leader Election, Log Replication, and Safety Guarantee. This design approach has led to its widespread adoption in many practical systems, such as etcd \cite{b27}, CockroachDB \cite{b36}, TiDB \cite{b37}, and PolarFS \cite{b38}.
\par
\begin{figure}[!tbp]
\centerline{\includegraphics[width=0.43\textwidth]{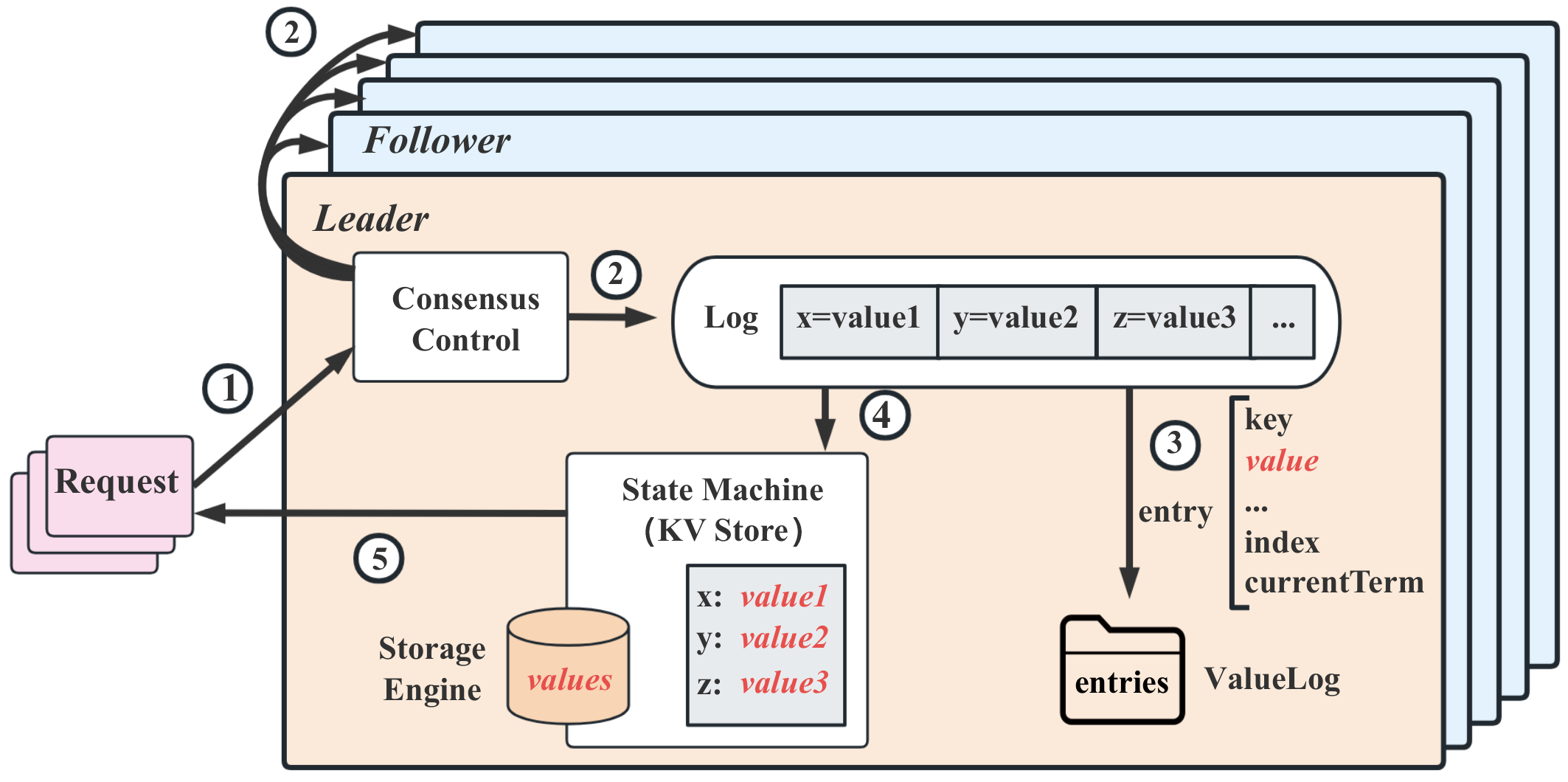}}
    \caption{An example of the request processing of Raft. 
    %The red text elements related to ``value" represent data that requires persistent storage operations.
    }
    \label{FIG1}
    \vspace{-2em}
\end{figure}

Raft achieves consensus with a leader-based approach. One node is elected as the leader and takes full responsibility for managing the replicated log. Other nodes act as followers, as shown in Figure~\ref{FIG1}. In the standard implementation of Raft, any modification to the state of the system must undergo a rigorous protocol process: the Leader node first receives client requests (Step \textcircled{1}), appends the requested content as new Raft log entries to its local log, and simultaneously synchronizes them to followers in the cluster through the log replication mechanism (Step \textcircled{2}). To ensure the safety properties of the algorithm, Raft enforces the following key constraints:
\par

\begin{itemize}
    \item \textit{Log Persistence}: All log entries must be persistently written to the ValueLog (a dedicated disk file for persisting Raft log entries)  to prevent data loss in the event of node crashes (Step \textcircled{3}).
    \item \textit{Majority Confirmation}: Log entries can only be committed to the state machine after being confirmed by a majority of nodes in the cluster (Steps \textcircled{4}-\textcircled{5}).
    \item \textit{State Recovery}: Nodes must be able to fully recover committed logs from persistent storage after restart.
\end{itemize}
\par
These safety requirements lead to frequent data persistence operations, as shown by the ``value" in Figure~\ref{FIG1}. Particularly in distributed key-value storage systems, these operations interweave with the storage engine's own persistence mechanisms, causing additional performance overhead.

\subsection{Key-Value Separation}
\label{section2.3}
Key-value separation is a storage optimization technique designed to reduce write amplification in LSM-tree-based systems. WiscKey \cite{b12} pioneered this architecture, demonstrating substantial performance improvements over traditional LSM-tree designs. The core idea is to store only keys and value pointers in the LSM-tree while placing actual values in a separate append-only log file. During compaction, only the lightweight keys and pointers are rewritten, avoiding the repeated movement of large values across LSM-tree levels. This design significantly reduces write amplification and also decreases the LSM-tree size, leading to better write performance and caching efficiency.

\subsection{Motivation}
\label{section2.4}
The co-design of consensus protocols and storage engines has recently emerged as an important research direction in distributed storage systems. PASV~\cite{b50} identified the double-logging problem between Raft and LSM-tree, proposing to eliminate the storage-layer WAL. PALF~\cite{b49} redesigned replicated write-ahead logging for distributed databases. More recently, LSM-Raft~\cite{b76} optimizes follower-side redundant writes by transmitting compacted Sorted String Table (SSTable) files instead of fine-grained log entries. These recent works indicate that the redundant persistence between the consensus layer and storage layer has become a critical performance bottleneck.

Meanwhile, recent hardware evolution has further amplified the urgency of this problem. The proliferation of high-performance NVMe SSDs makes software-level I/O inefficiencies the new bottleneck. Furthermore, in cloud-native environments with usage-based storage billing, eliminating redundant writes directly translates to cost savings.

\begin{figure}[tbp]
    \centerline{\includegraphics[width=0.43\textwidth]{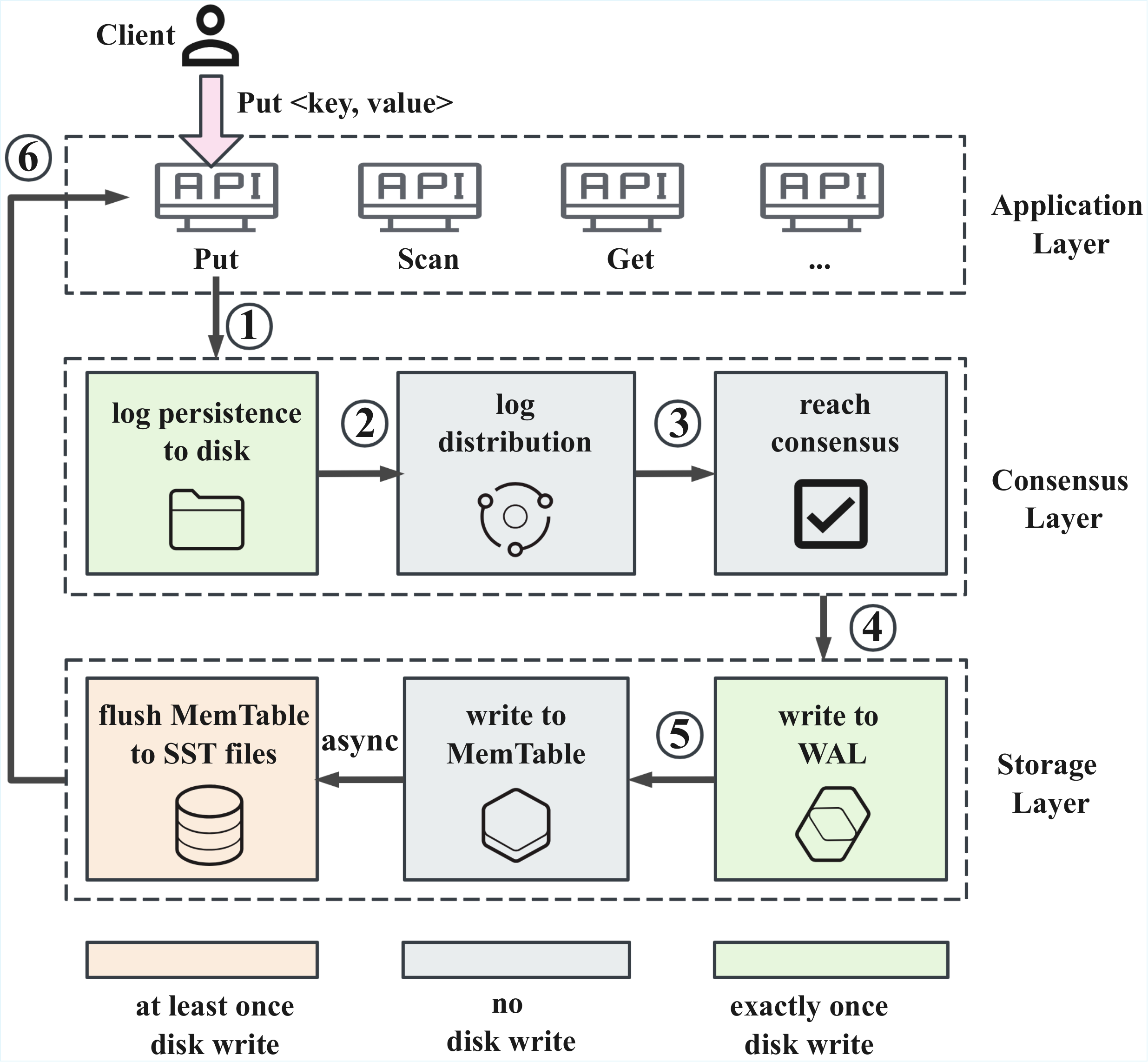}}
    \caption{Processing flow of write operations in a Raft-based distributed key-value store with LSM-tree storage engine.}
    \label{FIG2}
    % \vspace{-2em}
\end{figure}

\textbf{\textit{Analysis of Redundant Persistence Operations}}. 
Motivated by these trends, we conduct an in-depth analysis of the redundant persistence problem in existing Raft-based distributed key-value storage systems. The Raft protocol requires each request to be persistently stored as log entries to preserve the consensus state and enable fast recovery and synchronization after node restarts. Meanwhile, storage engines implement their own persistence mechanisms to ensure data reliability and support failure recovery. As shown in Figure~\ref{FIG2}, in distributed key-value storage systems based on Raft and LSM-tree, each write request must undergo the following critical persistence operations: 1) serializing and writing request content to persistent Raft logs; 2) writing to LSM-tree's WAL after consensus is reached; 3) flushing data to Sorted String Table (SST, SSTable) files when the MemTable reaches its capacity threshold. Note that this analysis focuses on LSM-tree-based distributed storage systems, and we do not extend the discussion to other storage architectures in this work.
\par

% \textbf{\textit{Opportunity for Integration of Key-Value Separation with Raft}}.
% To address the redundant data persistence issues in the system, we discovered a unique optimization opportunity by integrating key-value separation technology with the Raft protocol. As a widely adopted optimization strategy in storage systems \cite{b13,b14,b15}, key-value separation has demonstrated significant effectiveness in improving storage performance. Research work represented by Wisckey \cite{b12} has fully illustrated the potential of this technology. Its core concept is to decouple the storage of key-value pairs: storing keys and position pointers (substantially smaller than actual values) in the storage engine while maintaining actual values in separate log files. This design significantly reduces the volume of data managed by the storage engine, effectively mitigating write amplification issues.
\textit{\textbf{Opportunity for Integration of Key-Value Separation with Raft}}. 
To address the redundant data persistence issues in the system,
we discovered a unique optimization opportunity by integrating key-value separation 
% \sout{ technology with the Raft protocol. As a widely adopted optimization strategy in storage systems \cite{b13,b14,b15}, key-value separation has demonstrated significant effectiveness in improving storage performance. Research work represented by Wisckey \cite{b12} has fully illustrated the potential of this technology. Its core concept is to decouple the storage of key-value pairs: storing keys and position pointers (substantially smaller than actual values) in the storage engine while maintaining actual values in separate log files. This design significantly reduces the volume of data managed by the storage engine, effectively mitigating write amplification issues.}
(introduced in Section \ref{section2.3}) with the Raft protocol. By storing only lightweight offsets in the storage engine while maintaining values in Raft logs, we can unify value persistence with log replication, thereby eliminating redundant persistence operations at the storage layer.
\par
Through innovative integration of key-value separation with Raft's log replication mechanism, we propose a novel storage architecture. In this design, values are only stored in Raft logs, unifying value persistence with log replication and thereby eliminating redundant persistence operations at the storage layer. 
% Through innovative integration of key-value separation with Raft's log replication mechanism, we propose a novel storage architecture. In this design, values are stored exclusively in Raft logs as individual log entries in a disk file called ValueLog, and returns the storage offset of the written entry. When the corresponding Raft log reaches consensus, only the key and its offset are applied to the storage layer. Thus, the storage layer's WAL and Memtable contain only lightweight offsets rather than complete values, effectively eliminating value duplication between consensus and storage layers.
However, this innovative architecture introduces two key technical challenges:
\par
\begin{itemize}
    \item \textit{GC under Safety Guarantees}: Raft ensures data consistency by maintaining log sequentiality and completeness, but
    current GC mechanisms in key-value storage systems may
    compromise these fundamental properties.
    \item \textit{Read Performance Optimization}: The inherent characteristics of key-value separation necessitate additional I/O overhead for reconstructing key-value associations when processing read requests (Get and Scan operations), potentially leading to significant degradation in read performance.
\end{itemize}
\par
To address these challenges, we design a Raft-aware GC framework that achieves a dynamic balance between read and write performance. This framework precisely controls GC triggering timing, reorganizes scattered valid data into sequential disk files, and constructs efficient indexing structures to accelerate data access. Specifically, we optimize performance for both point and range queries: maintaining a hash index for key-to-offset mapping accelerates point queries, while the sequential organization of data enhances range query efficiency. This solution not only mitigates the read performance degradation caused by key-value separation but also ensures consistency and reliability of the GC process through deep integration with Raft's log management mechanism. 
Ultimately, our approach maintains the write efficiency advantages of key-value separation while effectively addressing its inherent limitations in read performance.

\section{NEZHA}
\label{sectio3}
Based on our analysis of the interaction between consensus protocols and storage engines in distributed key-value systems (Section \ref{sectio2}), we present Nezha, a novel distributed key-value storage system. 
The system achieves high performance through three key architectural innovations: 1) a storage-optimized consensus mechanism that significantly reduces write I/O overhead by persisting values and Raft-related metadata only once in ValueLog; 2) a systematic GC framework that substantially improves read performance while optimizing storage efficiency; 3) a three-phase request processing mechanism that ensures correct and efficient request handling during the entire storage lifecycle.
Detailed designs of these system components are presented in Sections \ref{section3.2}, \ref{storage management}, and \ref{Three-Phase}, respectively. Section \ref{Fault Tolerance} discusses Nezha's fault tolerance and safety guarantees.

\subsection{Overview}

\begin{figure}[tbp]
    \centerline{\includegraphics[width=0.46\textwidth]{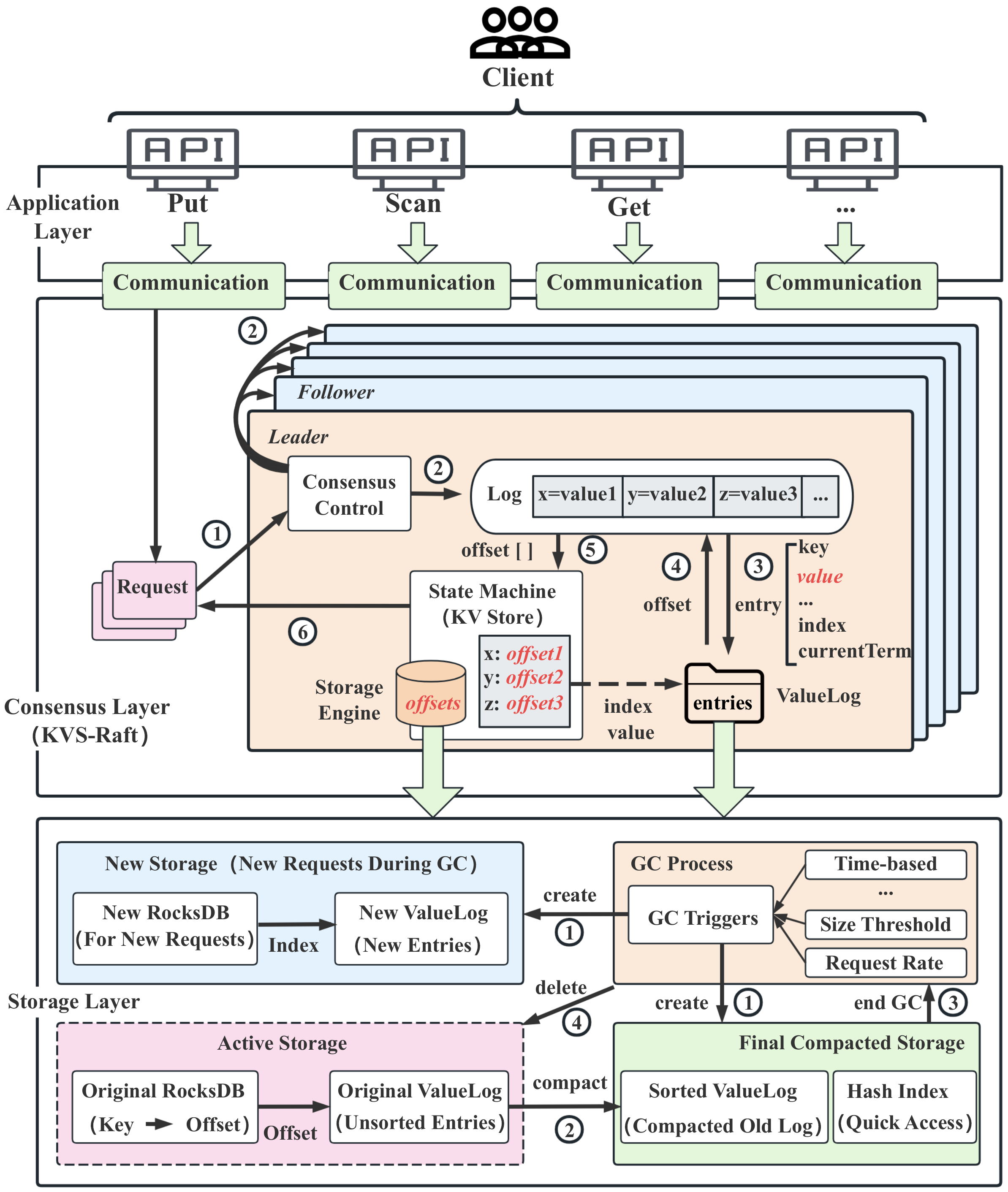}}
    \caption{The architecture of Nezha.}
    \label{FIG3}
    \vspace{-2em}
\end{figure}

% As illustrated in Figure~\ref{FIG3}, Nezha adopts a layered architectural design comprising three core components: the Application Layer, Consensus Layer, and Storage Layer.
Modern distributed key-value storage systems, such as TiKV \cite{b48}, can be conceptually organized into three fundamental layers: Application, Consensus, and Storage. As illustrated in Figure~\ref{FIG3}, Nezha follows this layered architecture.
\textit{\textbf{At the application layer}}, the system provides comprehensive key-value storage interfaces, including Put, Get, and Scan operations, which not only satisfy diverse data storage requirements but also ensure compatibility with existing systems.
% \textit{\textbf{At the application layer}}, the system provides standard key-value operations (Put, Get, and Scan) that support diverse storage requirements while maintaining compatibility with existing systems.
% \par
\textit{\textbf{At the consensus layer}}, we propose the KVS-Raft algorithm, which innovatively integrates the key-value separation strategy into the Raft consensus protocol to optimize system performance. 
\textit{\textbf{At the storage layer}}, we design a storage management mechanism deeply adapted to KVS-Raft. Our core contribution is a designed GC framework that divides request processing into three distinct phases: Pre-GC, During-GC, and Post-GC. This fine-grained phase division ensures correct handling of concurrent user requests during GC, maintaining strong consistency while guaranteeing system availability.
% \textit{\textbf{At the storage layer}}, we design a GC framework specifically tailored for KVS-Raft that optimizes storage while maintaining system correctness. To ensure uninterrupted service during storage optimization, this framework divides request processing into three distinct phases: Pre-GC, During-GC, and Post-GC. The GC timing and state transitions are transparent to the consensus layer and do not affect data consistency guarantees. When the consensus layer applies committed requests to the state machine, the GC framework handles each request according to its current phase, ensuring the state machine presents consistent and correct states to external clients.
\par

Through deep integration of key-value separation mechanisms at both consensus and storage layers, Nezha not only strictly guarantees strong consistency in distributed environments but also achieves significant performance optimization. This cross-layer collaborative optimization strategy fully leverages the advantages of key-value separation while effectively mitigating its potential limitations.
% By extending key-value separation from the traditional storage layer to the higher-level consensus layer, Nezha enables the benefits of key-value separation to span across both consensus and storage layers rather than being confined to storage alone. Additionally, we design an adaptive GC framework specifically tailored to address the challenges introduced by integrating key-value separation into KVS-Raft. This cross-layer collaborative optimization strategy fully leverages the advantages of key-value separation while effectively mitigating its potential limitations.

\subsection{Optimized Raft Log Storage Mechanism}
\label{section3.2}
In traditional distributed key-value storage systems, consensus algorithms face significant performance bottlenecks in write request processing. The root cause lies in the requirement of multiple disk write operations for large-volume value data to ensure consistency and durability, which leads to excessive I/O overhead and significantly increased request processing latency. To address this challenge, we propose an innovative approach that integrates the Raft protocol with the key-value separation strategy.
% In traditional distributed key-value storage systems, Raft requires persisting complete key-value pairs as log entries to ensure consensus correctness and enable recovery after node failures. This persistence is essential for maintaining data consistency across the distributed cluster. However, when these committed log entries are subsequently applied to the storage layer's state machine, the large values undergo redundant writes to WAL and Memtable, resulting in unnecessary I/O overhead for the same data. To address this redundancy, we propose an innovative approach that integrates key-value separation with the Raft protocol. Our design keeps values in the consensus layer's ValueLog while allowing only lightweight offsets to accompany keys into the storage layer, thereby eliminating redundant writes of large values.
\par
% As depicted in the \textbf{consensus layer} of Figure~\ref{FIG3}, the write request processing flow of KVS-Raft implements an optimized mechanism to achieve efficient data persistence:
Specifically, as depicted in the \textbf{consensus layer} of Figure~\ref{FIG3}, KVS-Raft achieves this optimization through the following write request processing flow: 
First, client write requests initially arrive at the Consensus Control module of the leader node (Step \textcircled{1}). This module encapsulates write requests into Raft log entries, performing local log appending while simultaneously synchronizing them to followers (Step \textcircled{2}).
Once the Raft log is appended, the system serializes the key-value pair and its consensus-related metadata (such as \textit{currentTerm} and \textit{index}) as an \textit{entry} entity, persists it to the ValueLog file (Step \textcircled{3}), and subsequently obtains the corresponding file offset for this \textit{entry} (Step \textcircled{4}).
% Once the Raft log is appended, the system stores the key, value, and essential recovery metadata (such as \textit{currentTerm} and \textit{index}) as an \textit{entry} entity in the ValueLog file (Step \textcircled{3}), and returns the corresponding file offset for this \textit{entry} (Step \textcircled{4}).
Subsequently, the leader node only needs to \textit{apply the lightweight offset (rather than the large-volume value)} to its state machine (Step \textcircled{5}). Finally, the system returns a response to the client, completing the write operation (Step \textcircled{6}).
\par
This optimized storage architecture provides several key benefits.
First, by persisting value data only once to ValueLog and maintaining only lightweight offsets in the state machine, the system \textit{reduces the number of disk writes for value data from at least three times in traditional approaches to just once}, significantly reducing I/O overhead.
% Second, the lightweight state machine design (storing only offsets instead of complete values) not only reduces storage space consumption but also accelerates state updates and recovery processes.
Second, the lightweight state machine design (storing only offsets instead of complete values) not only reduces storage space consumption but also accelerates state updates and recovery processes, as demonstrated in our experimental evaluation (Section \ref{section4.7}).
More importantly, this integration of key-value separation into the Raft consensus layer achieves enhanced write performance while fully preserving Raft's safety properties, providing a practical solution for building high-performance distributed key-value storage systems.

\subsection{Adaptive Storage Management Mechanism}
\label{storage management}
While the optimized storage mechanism for Raft logs significantly enhances write performance, it introduces two key technical challenges. First, the inherent conflict between Raft's requirement for strict log ordering (essential for consensus agreement and state replication) and the non-sequential nature of key-value separation storage renders traditional GC \cite{b12} unsuitable, as they might compromise the integrity and sequentiality of Raft logs. Second, key-value separation introduces additional I/O overhead when processing read requests, as reconstructing key-value associations requires extra disk operations. To address these challenges, we design an adaptive storage management framework, as illustrated in \textbf{the storage layer} of Figure~\ref{FIG3}. This framework encompasses two core types of storage files:

\begin{itemize}
    \item \textit{Unordered Log Files:} These files store initial Raft log entries in append-only chronological order, maintaining complete log information to support Raft's replication and recovery mechanisms.
    \item \textit{Ordered Log Files:} These files store sorted and compacted Raft log entries, implementing key-based sequential storage to optimize range query performance while eliminating redundant and expired data.
\end{itemize}
% \begin{table}[tbp]    % 使用普通table环境，在单栏内显示
% \centering            % 在当前栏内居中
% \caption{Storage module usage for read and write operations across GC phases.}
% \label{tabel1}
% \renewcommand{\arraystretch}{1.0}
% \begin{tabular}{ccc}   % 改为 ccc，添加第三列
% \hline
% \rule{0pt}{1.5ex}
% \textbf{Phase} & \textbf{Read} & \textbf{Write} \\[0.5ex]
% \hline
% Pre-GC & Active & Active \\
% During-GC & New + Active & New \\
% Post-GC & New + Final Compacted & New \\
% \hline
% \end{tabular}
% \end{table}

The \textbf{\textit{Active Storage Module}} serves as the primary write receiver, handling real-time read and write requests and maintaining unordered log files. Benefiting from the KVS-Raft optimization, this module only needs to store lightweight offsets rather than complete values in RocksDB \cite{b20} (a popular single-machine LSM tree-based storage engine), enabling efficient write support. To manage the accumulated log files effectively, the system initiates GC based on multidimensional triggers, including storage space thresholds, scheduled timing mechanisms, and request load levels. While this comprehensive triggering mechanism ensures efficient storage utilization and stable system performance, the system must maintain read and write availability during GC, thus necessitating additional storage components.
\par
The \textbf{\textit{New Storage Module}} is specifically designed to handle read and write requests during GC. 
Similar to the Active Storage, it consists of a ValueLog file for storing new entries and a RocksDB instance for maintaining lightweight key-offset mappings. 
When the system initiates the GC, all new read and write requests are redirected to this module, ensuring continuous system availability by maintaining independent index structures. This design enables GC to proceed without compromising system availability while providing clear boundaries for eventual data consolidation.
\par
The \textbf{\textit{Final Compacted Storage Module}}, as the output of GC, stores reorganized and compacted ordered data. This module significantly enhances range query performance through a combination of hash index structures and the inherent sequential layout of sorted data. Notably, the sorted ValueLog reflects the state machine status corresponding to the original ValueLog in the Active Storage. \textit{\textbf{Our system creates snapshots by integrating the sorted ValueLog with two variables from the original ValueLog: the \textit{last index} and 
the \textit{last term}, which aligns with the log compaction mechanism described in the Raft paper \cite{b5}.}} Consequently, upon completion of GC, the system can safely remove the old ValueLog and corresponding RocksDB files from the Active Storage. At this point, the Final Compacted Storage not only reduces storage space consumption through efficient data layout and compression strategies but also ensures crash consistency and system state integrity via its snapshot-based recovery mechanism.

As depicted in the \textbf{storage layer} of Figure~\ref{FIG3}, the system employs different combinations of storage modules across various phases. This systematic modular architecture provides clear data boundaries, enabling precise targeting of appropriate storage components for read and write operations. To systematically analyze the GC framework, we decompose its lifecycle into the following key phases: (1) \textit{GC Initialization:} The system creates a New Storage to handle incoming write requests while initializing the sorted ValueLog within Final Compacted Storage for storing ordered data entries (Step \textcircled{1}); (2) \textit{Data Compaction:} The system reorganizes ValueLog from the Active Storage through compression and sorting operations, generating an ordered ValueLog in the Final Compacted Storage. Concurrently, all read and write requests are redirected to the New Storage (Step \textcircled{2}); (3) \textit{Cleanup Phase:} Upon GC completion, the system safely eliminates expired files from the Active Storage (Steps \textcircled{3}-\textcircled{4}); (4) \textit{Steady State:} The system enters a stable state, continuously monitoring storage utilization of New Storage and other critical metrics.
% 表1
\begin{table}[tbp]    % 使用普通table环境，在单栏内显示
\centering            % 在当前栏内居中
\caption{Combinations of storage modules at different system phases.}
\label{tabel1}
\renewcommand{\arraystretch}{1}
\begin{tabular}{cc}   % 将 ll 改为 cc，使两列都居中对齐
\hline
\rule{0pt}{1.2ex}
\textbf{Phase} & \textbf{Storage Module} \\[1ex]
\hline
Pre-GC & Active Storage \\
During-GC & New Storage, Active Storage \\
Post-GC & New Storage, Final Compacted Storage \\
\hline
\end{tabular}
\vspace{-2em}
\end{table}

Notably, this mechanism operates in an iterative cycle: when New Storage reaches predetermined thresholds or meets other triggering conditions, the system initiates a new round of GC. In subsequent GC cycles, the current New Storage transitions to become the compaction target, with its data being systematically merged into the existing Sorted ValueLog while a New Storage is established to process incoming read and write requests. This cyclical design ensures continuous optimization of system storage efficiency while maintaining service continuity. More details on request handling mechanisms are discussed in Section \ref{Three-Phase}.

% {\color{R5} \textbf{[R5-O1]} \textbf{\textit{Phase Transition Mechanism.}} To clarify the complete GC lifecycle, we describe how the system transitions from the Post-GC phase of the first cycle to the Pre-GC phase of the second cycle. Upon GC completion, the following sequence occurs: (1) the Final Compacted Storage becomes fully operational with sorted ValueLog and hash index ready for historical queries; (2) the system safely removes the obsolete Active Storage files and updates the GC state flags; (3) the New Storage from the first cycle logically becomes the Active Storage and continues handling read and write requests, while the Final Compacted Storage from the first cycle logically becomes the Last Compacted Storage and serves historical data lookups. This state corresponds to the Pre-GC phase of the second cycle. When the Active Storage accumulates sufficient data and reaches triggering thresholds, the system initiates the second GC cycle, creating both a fresh New Storage and a new Final Compacted Storage to handle incoming write requests and store the newly compacted data, respectively. Note that Table~\ref{tabel1} describes the first GC cycle; in subsequent cycles, the Pre-GC phase involves both Active Storage (for recent data) and Last Compacted Storage (for historical data) working collaboratively.}

\textbf{\textit{Phase Transition Mechanism.}} Upon GC completion, the system transitions from Post-GC to the Pre-GC phase of the next cycle through the following steps: (1) the system safely removes the obsolete Active Storage files; (2) the New Storage then logically becomes the Active Storage for the next cycle; (3) the GC state flags are reset to indicate the Pre-GC phase. This role rotation ensures continuous system availability across multiple GC cycles.

\subsection{Three-Phase Request Processing Mechanism}
\label{Three-Phase}
To ensure correct request handling throughout the complete GC lifecycle, we partition the request processing mechanism into three distinct phases: Pre-GC, During-GC, and Post-GC, as shown in Table \ref{tabel1}. This section focuses on the handling strategies for core operations (Put, Get, and Scan) across different phases.

It is worth noting that read and write operations require different handling strategies with respect to GC phases. Write operations (Put) are GC-phase-agnostic: they always write to the storage module currently referenced by the file descriptors (\textit{currentLog} and \textit{currentDB}), which are automatically switched when GC is triggered. In contrast, read operations (Get and Scan) must be GC-phase-aware, as they may need to query multiple storage modules with different lookup paths to retrieve the most recent data. Accordingly, we present a unified algorithm for Put, followed by phase-specific analyses for Get and Scan.

% 算法1
\begin{algorithm}[t]
\caption{Write Operation Handler}
\label{put}
\begin{algorithmic}[1]
\setlength{\baselineskip}{0.8\baselineskip}
\Statex \textbf{Input:} $(k,v)$: Key-value pair to write.
\Statex \textbf{Output:} Operation status: SUCCESS or TIMEOUT.
\Function{HandleWrite}{$k,v$}
\Statex \hspace{\algorithmicindent} \textbf{Phase 1: Log Entry Creation and Persistence}
\State $entry$ $\gets$ \Call{ConstructRaftLogEntry}{$k,v$}
\State $offset$ $\gets$ \Call{PersistToLog}{\textit{currentLog}, $entry$}
\Statex \hspace{\algorithmicindent} \textbf{Phase 2: Consensus and State Machine Application}
\State $t$ $\gets$ \Call{InitializeTimer}{\textit{CONSENSUS\_TIMEOUT}}
\State $result$ $\gets$ \Call{WaitForCommitOrTimeout}{$entry$, $t$}
\If{$result$ = \texttt{COMMITTED}}
    \State \Call{ApplyStateMachine}{\textit{currentDB}, $k$, $offset$}
    \State \Return \texttt{SUCCESS}
\Else
    \State \Return \texttt{TIMEOUT}
\EndIf
\EndFunction
\end{algorithmic}
\end{algorithm}

\par
\subsubsection{\textbf{Put Request Processing}}
Algorithm \ref{put} demonstrates an optimized write operation mechanism that integrates efficient log persistence with state machine application. The algorithm leverages two primary storage components: \textit{currentLog} for value storage and \textit{currentDB} for key-offset mappings, while employing a \textit{CONSENSUS\_TIMEOUT} to ensure operation liveness.
\par

The write operation consists of two distinct phases: (1) \textit{Log Entry Creation and Persistence (Lines 2-3):} The system first constructs a Raft log entry containing the key-value pair information. This entry is then synchronously persisted to disk in the current log file, generating an offset that uniquely identifies the value's location. In our optimized design, this critical step \textbf{\textit{serves as the one and only write operation for values}}; (2) \textit{Consensus and State Machine Application (Lines 4-10):} After log persistence, the system initiates a consensus process with a bounded timeout (Lines 4-5). Upon receiving commit confirmation through the Raft protocol (Line 6), the system applies the operation to its state machine by storing only the lightweight offset in the storage engine, rather than the entire value (Line 7). 
\par
This design achieves significant efficiency improvements through key-value separation: by persisting values only once during log entry creation and storing lightweight offsets in the state machine, the system \textbf{\textit{reduces the number of value writes from at least three to just one}}. This optimization is particularly effective for large-value operations while maintaining Raft's strong consistency guarantees.

\par
\subsubsection{\textbf{Get Request Processing}}
Algorithm \ref{get} illustrates the point query processing strategy across different phases. The implementation relies on several essential state variables: \textit{GC\_Started} and \textit{GC\_Completed}, boolean flags indicating GC phase transitions; \textit{oldDB} and \textit{newDB}, RocksDB instances managing key-offset mappings for original and new data respectively; and \textit{oldLog} and \textit{currentLog}, representing paths to ValueLog files for unsorted original entries and newly written entries; and \textit{sortedFile}, representing the value-sorted file containing compacted historical data after GC completion.

% 算法2
\begin{algorithm}[t]
\caption{Parallel Point Query Handler}
\label{get}
\begin{algorithmic}[1]
\setlength{\baselineskip}{0.8\baselineskip}
\Statex \textbf{Input:} $k$: Query key.
\Statex \textbf{Output:} Value associated with $k$ or NOT\_FOUND.
\Function{HandleRead}{$k$}
\Statex \hspace{\algorithmicindent} \textbf{Phase 1: Pre-GC State}
\If{$\neg$ \texttt{GC\_Started}}
    \State $offset$ $\gets$ \Call{QueryOffset}{\textit{oldDB}, $k$}
    \If{$offset = -1$} \Return \texttt{NOT\_FOUND} \EndIf
    \State \Return \Call{ReadValue}{\textit{oldLog}, $offset$}
\EndIf
\Statex \hspace{\algorithmicindent} \textbf{Phase 2: During-GC State}
\If{$\neg$ \texttt{GC\_Completed}}
    \State $f_{new\_offset}$ $\gets$ \Call{AsyncQueryOffset}{\textit{newDB}, $k$}
    \State $f_{old\_offset}$ $\gets$ \Call{AsyncQueryOffset}{\textit{oldDB}, $k$}
    \State $offset$ $\gets$ \Call{Await}{$f_{new\_offset}$}
    \If{$offset \neq -1$}
        \State \Return \Call{ReadValue}{\textit{currentLog}, $offset$}
    \EndIf
    \State $offset$ $\gets$ \Call{Await}{$f_{old\_offset}$}
    \If{$offset = -1$} \Return \texttt{NOT\_FOUND} \EndIf
    \State \Return \Call{ReadValue}{\textit{oldLog}, $offset$}
\EndIf
\Statex \hspace{\algorithmicindent} \textbf{Phase 3: Post-GC State}
\State $f_{new\_offset}$ $\gets$ \Call{AsyncQueryOffset}{\textit{newDB}, $k$}
\State $f_{sorted\_value}$ $\gets$ \Call{AsyncQueryValue}{\textit{sortedFile}, $k$}
\State $offset$ $\gets$ \Call{Await}{$f_{new\_offset}$}
\If{$offset \neq -1$}
    \State \Return \Call{ReadValue}{\textit{currentLog}, $offset$}
\EndIf
\State $value$ $\gets$ \Call{Await}{$f_{sorted\_value}$}
\If{$value = -1$} \Return \texttt{NOT\_FOUND} \EndIf
\State \Return $value$
\EndFunction
\end{algorithmic}
\end{algorithm}

The point query processing mechanism consists of three distinct phases:
\par
\begin{itemize}
    \item \textit{Pre-GC Phase (Lines 2-7):} Before GC starts, the system exclusively queries the Active Storage, first retrieving the \textit{offset} from the storage engine file referenced by \textit{oldDB} (Line 3). If no \textit{offset} is found in the storage engine (i.e., \textit{offset} equals -1), the key is determined to be \textit{NOT\_FOUND} (Line 4); otherwise, the corresponding \textit{value} is retrieved from the log file referenced by \textit{oldLog} (Line 6).
    
    \item \textit{During-GC Phase (Lines 8-19):} The system implements parallel lookups to optimize query efficiency. It simultaneously initiates two asynchronous queries for key-offset mappings: one to New Storage and another to Active Storage (Lines 9-10). Upon receiving the offset from New Storage, if it exists, the system immediately retrieves the corresponding value from \textit{currentLog} (Lines 11-13)
    % , which is illustrated by the red dotted line in Figure~\ref{query}
    . This prioritization ensures access to the most recent data version. Only when the key is not found in New Storage does the system utilize the offset from Active Storage to fetch the value from \textit{oldLog} (Lines 15-18).
    
    \item \textit{Post-GC Phase (Lines 20-29):} The system simultaneously initiates an asynchronous offset query to New Storage and a direct value query to the \textit{sortedFile} (Lines 20-21). If an offset is found in New Storage, the system immediately retrieves the value from \textit{currentLog} (Lines 22-24), as this represents the most recent data. Otherwise, it returns the value obtained directly from the \textit{sortedFile}, which contains all historical data in an optimized format (Lines 26-28). %This dual-path query mechanism ensures optimal performance while maintaining consistency.
\end{itemize}
% \par
This systematic three-phase design ensures continuous read service availability during the storage structure optimization process. Through strategically designed lookup sequences, the system maintains both strong consistency and efficient query performance.

% 算法3
\begin{algorithm}[t]
\caption{Parallel Range Query Handler}
\label{scan}
\begin{algorithmic}[1]
\setlength{\baselineskip}{0.8\baselineskip}
\Statex \textbf{Input:} $[k_{start}, k_{end}]$: Query key range.
\Statex \textbf{Output:} Map of key-value pairs within range.
\Statex \textbf{Note:} State variables follow same semantics as Algorithm \ref{get}.
\Function{HandleRangeScan}{$k_{start}$, $k_{end}$}
\Statex \hspace{\algorithmicindent} \textbf{Phase 1: Pre-GC State}
\If{$\neg$ \texttt{GC\_Started}}
    \State \Return \Call{ScanKeys}{\textit{oldDB}, \textit{oldLog}, $k_{start}$, $k_{end}$}
\EndIf
\Statex \hspace{\algorithmicindent} \textbf{Phase 2: During-GC State}
\If{$\neg$ \texttt{GC\_Completed}}
    \State $f_{old}$ $\gets$ \Call{AsyncScan}{\textit{oldDB}, \textit{oldLog}, $k_{start}$, $k_{end}$}
    \State $f_{new}$ $\gets$ \Call{AsyncScan}{\textit{newDB}, \textit{currentLog}, $k_{start}$, $k_{end}$}
\Statex \hspace{\algorithmicindent} \textbf{Phase 3: Post-GC State}
\Else
    \State $f_{sorted}$ $\gets$ \Call{AsyncScanSorted}{\textit{sortedFile}, $k_{start}$, $k_{end}$}
    \State $f_{new}$ $\gets$ \Call{AsyncScan}{\textit{newDB}, \textit{currentLog}, $k_{start}$, $k_{end}$}
\EndIf
\State \Call{AwaitAll}{$f_{sorted}$ or $f_{old}$, $f_{new}$}
\State \Return \Call{MergeResults}{$f_{sorted}$ or $f_{old}$, $f_{new}$}
\EndFunction
\end{algorithmic}
\end{algorithm}

\subsubsection{\textbf{Scan Request Processing}}
Algorithm \ref{scan} illustrates the range query processing methodology across different phases. The mechanism uses \textit{sortedFile} as a key component, which maintains sorted data entries after GC and supports Raft log recovery, as described in Section \ref{storage management}. The range query processing involves three different scenarios based on the GC phase:
\par
\begin{itemize}
    \item \textit{Pre-GC Phase (Lines 2-4):} Since GC has not started, the system exclusively queries data from the Active Storage (Line 3).
    \item \textit{During-GC Phase (Lines 5-7):} During GC, the system performs parallel queries in both Active Storage and New Storage modules (Lines 5-6), as new incoming requests are redirected to New Storage while existing data remains in Active Storage.
    \item \textit{Post-GC Phase (Lines 8-11):} After GC completion, the system performs parallel queries in both Final Compacted Storage and New Storage modules (Lines 9-10), accessing sorted historical data and newly written data, respectively.
\end{itemize}
\par
After completing all parallel queries, the system merges their results using a versioning-based strategy. During this consolidation, data from the New Storage (represented by {$f_{new}$}) takes precedence over entries from old or sorted files, thus maintaining the latest updates (Lines 12-13). This three-phase parallel query design, driven by a GC state, not only guarantees the correctness of range queries during system state transitions but also enhances query efficiency through parallel processing.

\subsection{Fault Tolerance and Safety Guarantees}
\label{Fault Tolerance}

%\subsubsection{Fault Model and Assumptions}
\textbf{\textit{Fault Tolerance}}.
Nezha adopts the standard Raft fault model. The system can tolerate up to $\left\lfloor \tfrac{n-1}{2} \right\rfloor$ node failures in an n-node cluster while maintaining safety and liveness properties. Nezha's node failure handling mechanisms are fundamentally identical to standard Raft, with differences arising only during garbage collection operations.

When failures occur during GC, the recovery process first checks the atomic GC state flag. If GC is incomplete, the system identifies the last key in the sorted file as the GC interrupt point and continues executing GC operations from that position, ensuring the integrity of GC operations.

\textbf{\textit{ValueLog Consistency and Recovery}}.
Cross-node ValueLog synchronization maintains consistency despite structural changes during GC. The system ensures that while physical ValueLog content may temporarily differ across nodes, the logical state machine remains consistent. Recovery leverages the sorted ValueLog from Final Compacted Storage as an efficient snapshot mechanism for follower catch-up.

\textbf{\textit{Formal Security Verification}}. We have conducted formal verification of Nezha's safety properties using TLA+ specifications, confirming that our modifications preserve Raft's correctness guarantees. The complete formal specifications and proofs are available in our open-source repository.
% \subsubsection{Failure Recovery Performance}

\section{IMPLEMENTATION AND EVALUATION}
\label{sectio4}
%In this section, we will introduce the implementation of Nezha in Section \ref{implementation}, and then present our experimental setup in Section \ref{section4.1}. The detailed evaluation results are then presented in Section \ref{section4.2} examining performance under different value sizes, Section \ref{Storage Distribution} analyzing the impact of data distribution between storage modules, and Section \ref{section4.4} evaluating system behavior under various Yahoo! Cloud Services Benchmark (YCSB) \cite{b6} workload patterns.

In this section, we will introduce the implementation of Nezha in Section \ref{implementation}, and then present our experimental setup in Section \ref{section4.2}. The detailed evaluation results are then presented in Section \ref{section4.3} examining performance under different value sizes, Section \ref{section4.4} evaluating the relationship between scan cardinality and range query efficiency, and Section \ref{section4.5} assessing system behavior under various Yahoo! Cloud Services Benchmark (YCSB) \cite{b6} workload patterns. 
Section \ref{scalability} examines system scalability across different cluster sizes,
Section \ref{section4.6} analyzes the impact of GC on system performance, and Section \ref{section4.7} evaluates recovery times under different GC states.

\subsection{System Implementation}
\label{implementation}
We implemented Nezha as a prototype system in Go (version 1.19.5).
The implementation leverages RocksDB as the underlying key-value storage engine, while the distributed communication framework is built using gRPC \cite{b7} in conjunction with Google Protocol Buffers \cite{b8} for efficient data serialization and message passing between nodes. 
All nodes are interconnected via a 10 Gigabit Ethernet (10GbE) network.
Our prototype contains around 7.5K lines of code (without gRPC and RocksDB).
The source code is available at https://github.com/Dshuishui/Nezha.

\subsection{Experimental Setup}
\label{section4.2}
\textbf{\textit{Environment and Parameters.}} The experiments were conducted on a cluster consisting of three high-I/O nodes, each equipped with Intel(R) Xeon(R) E5-2603 v3 (12 logical cores, 2.4 GHz), Ubuntu 20.04.4 LTS, 64 GB DRAM, and 2 TB SSD storage. 
%Three nodes served as Nezha nodes, while one simulated the YCSB benchmark tool as a load generation node.
\par

\textbf{\textit{Workloads.}} To systematically evaluate the performance benefits of Nezha's design, we conducted extensive comparative experiments across multiple dimensions with various workloads. For performance analysis based on value size, each key-value pair by default contained a 10 B key and values of varying sizes (1 KB to 256 KB), with each key-value pair having three replicas in Nezha. The key access pattern followed a Zipf distribution. We evaluated the performance of three basic operations in Nezha: Put, Get, and Scan. To better understand the system's behavior in practical scenarios, we also tested Nezha under YCSB's classic workload patterns that simulate real-world scenarios, as shown in Table \ref{table2}.
\par
\textbf{\textit{Baselines.}} To comprehensively evaluate Nezha's performance, we selected the following seven configurations for comparison:
\par
\begin{itemize}
    \item \textit{Original:} A traditional distributed key-value storage system that integrates the state-of-the-art Raft \cite{b3} consensus with the RocksDB storage engine.
     \item \textit{PASV \cite{b50}:} A solution for LSM-tree based relational databases that eliminates double-logging by removing the storage engine's WAL.
     \item \textit{TiKV \cite{b48}:} An enterprise-level distributed key-value storage system with an architecture similar to Original.
     \item \textit{Dwisckey:} A distributed implementation of Wisckey \cite{b12} that inherits the key-value separation design philosophy but extends it to a distributed environment.
     \item \textit{LSM-Raft \cite{b76}:} A recent consensus-storage co-design approach that transmits compacted SSTables instead of fine-grained log entries, reducing follower-side redundant compaction. However, leaders still experience full redundant writes during consensus.
    \item \textit{Nezha-NoGC:} The basic version of Nezha that implements key-value separation within the Raft consensus module. 
    \item \textit{Nezha:} The complete Nezha system with GC mechanism tightly coupled with Raft's log management.
\end{itemize}
\par
%\textbf{\textit{Metrics.}} We used the following metrics to evaluate different phases of Nezha and other solutions:
%\par
\begin{table}[tbp]
\centering
\caption{YCSB workloads used in the evaluation.}
\label{table2}
\renewcommand{\arraystretch}{1}
\setlength{\tabcolsep}{2pt} % Adjust this value as needed (e.g., 3pt, 4pt)
\begin{tabular}{cccc}
\hline
\rule{0pt}{1.0ex}
\textbf{Workload} & \textbf{Write Type} & \textbf{Query Type} & \textbf{Category} \\[0.5ex]
\hline
\textit{Load} & Insert & / & Insert Only \\
\textit{A} & Update & Point Query & 50\%write 50\%read \\
\textit{B} & Update & Point Query & 5\%write 95\%read \\
\textit{C} & / & Point Query & Read Only \\
\textit{D} & Insert & Point Query & 5\%write 95\%read \\
\textit{E} & Insert & Range Query & 5\%write 95\%scan \\
\textit{F} & \textit{RMW} & Point Query & 50\%write 50\%read \\
\hline
\end{tabular}
\end{table}
%\begin{itemize}
%    \item \textit{Throughput:} Amount of data processed in requests per second.
%    \item \textit{Latency:} Average execution time for read and write requests.
% \vspace{-0.3em}\

\subsection{Performance Analysis Based on Value Size}
\label{section4.3}
To validate the performance improvement of key-value separation for write operations and the balancing effect of the Raft-adapted GC mechanism on read performance, we evaluated the performance of Put, Get, and Scan operations in terms of throughput and average latency. The tests fixed the key size at 10 B while incrementing value sizes from 1 KB to 256 KB to analyze the impact of value size on system performance. To evaluate basic operational performance, we loaded 100 GB of data into the system, setting a 40 GB threshold for the active storage module to trigger GC.
%In Section \ref{section4.3}, we further evaluated the impact of the key-value pair length in a single scan on scan performance.
\par

\begin{figure*}[!t]
    \centering
    % 第一组子图（Put请求）
    \begin{minipage}[t]{0.31\textwidth}
        \centering
        \subfigure[Throughput of \textbf{\textit{Put}}]{
            \includegraphics[width=\textwidth]{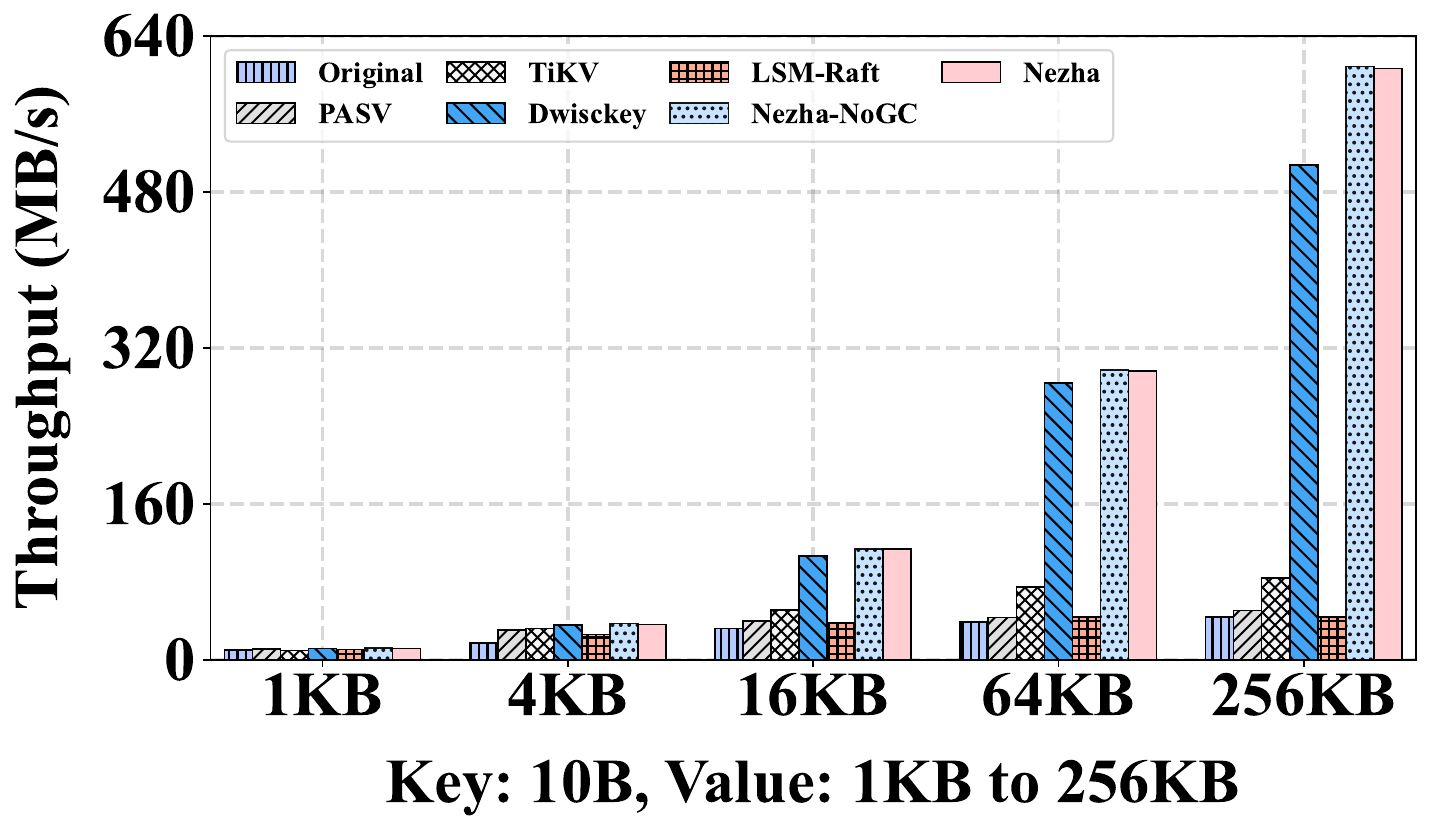}
            \label{put:throughput}
        }
        %\vspace{0.5em}
        \vspace{-0.1cm}
        \subfigure[Average write latency of \textit{\textbf{Put}}]{
            \includegraphics[width=\textwidth]{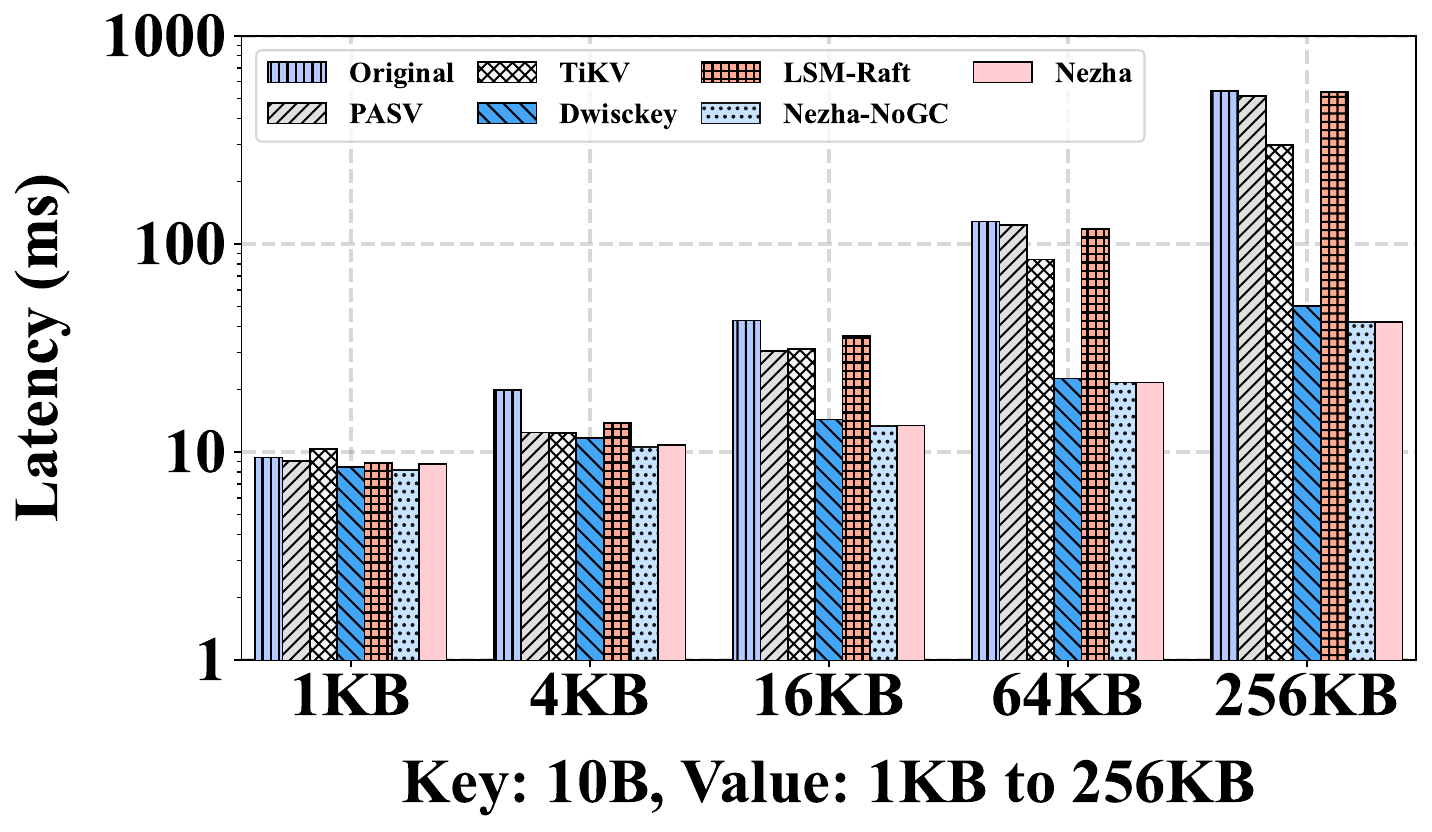}
            \label{put:latency}
        }
        %\vspace{-0.1cm}
        
        \caption{Performance comparisons for \textbf{put requests} under different value sizes.}
        \label{put-single}
    \end{minipage}
    \hfill
    % 第二组子图（Get请求）
    \begin{minipage}[t]{0.31\textwidth}
        \centering
        \subfigure[Throughput of \textit{\textbf{Get}}]{
            \includegraphics[width=\textwidth]{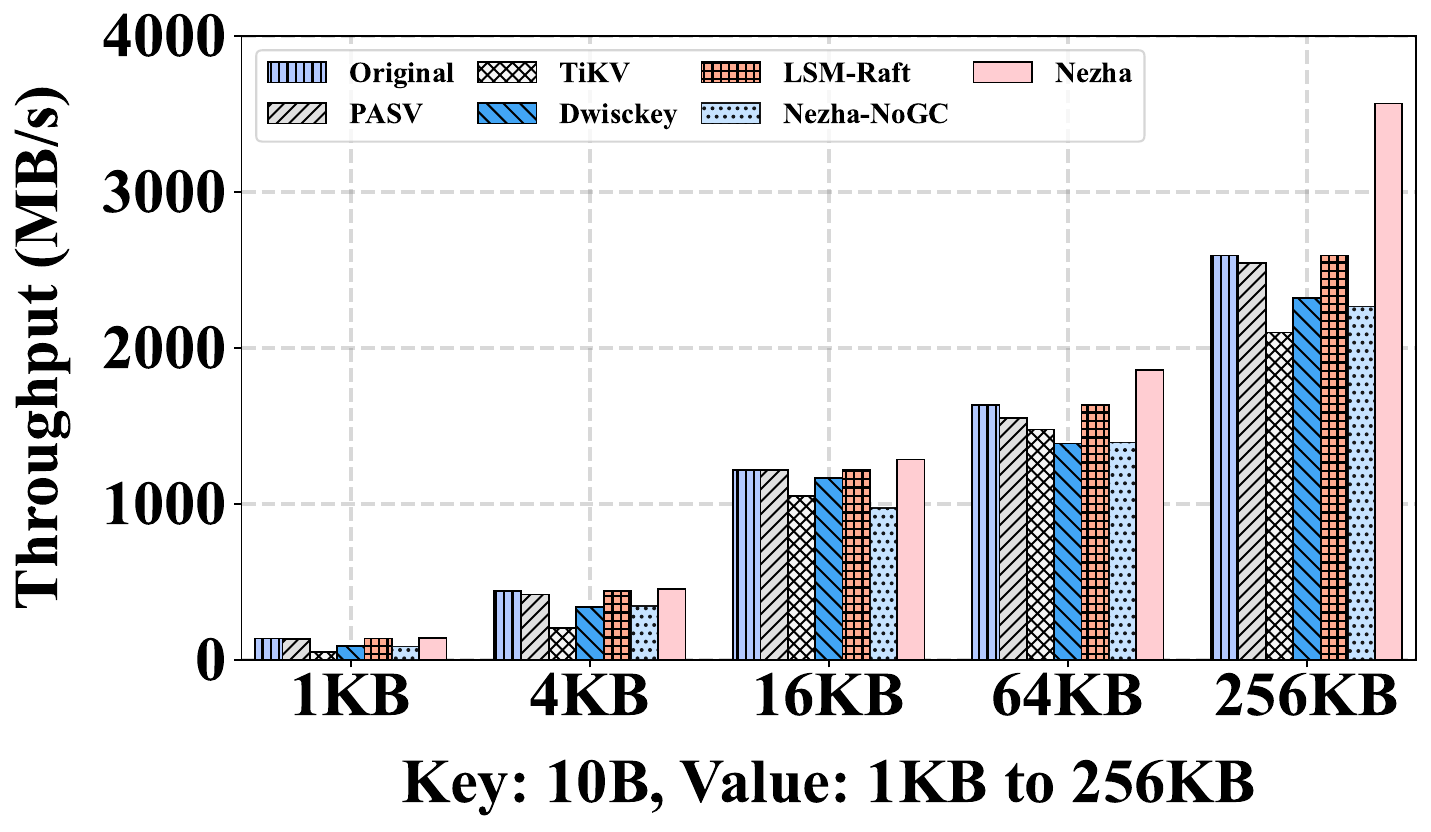}
            \label{get:throughput}
        }
        %\vspace{0.5em}
        \vspace{-0.1cm}
        \subfigure[Average read latency of \textbf{\textit{Get}}]{
            \includegraphics[width=\textwidth]{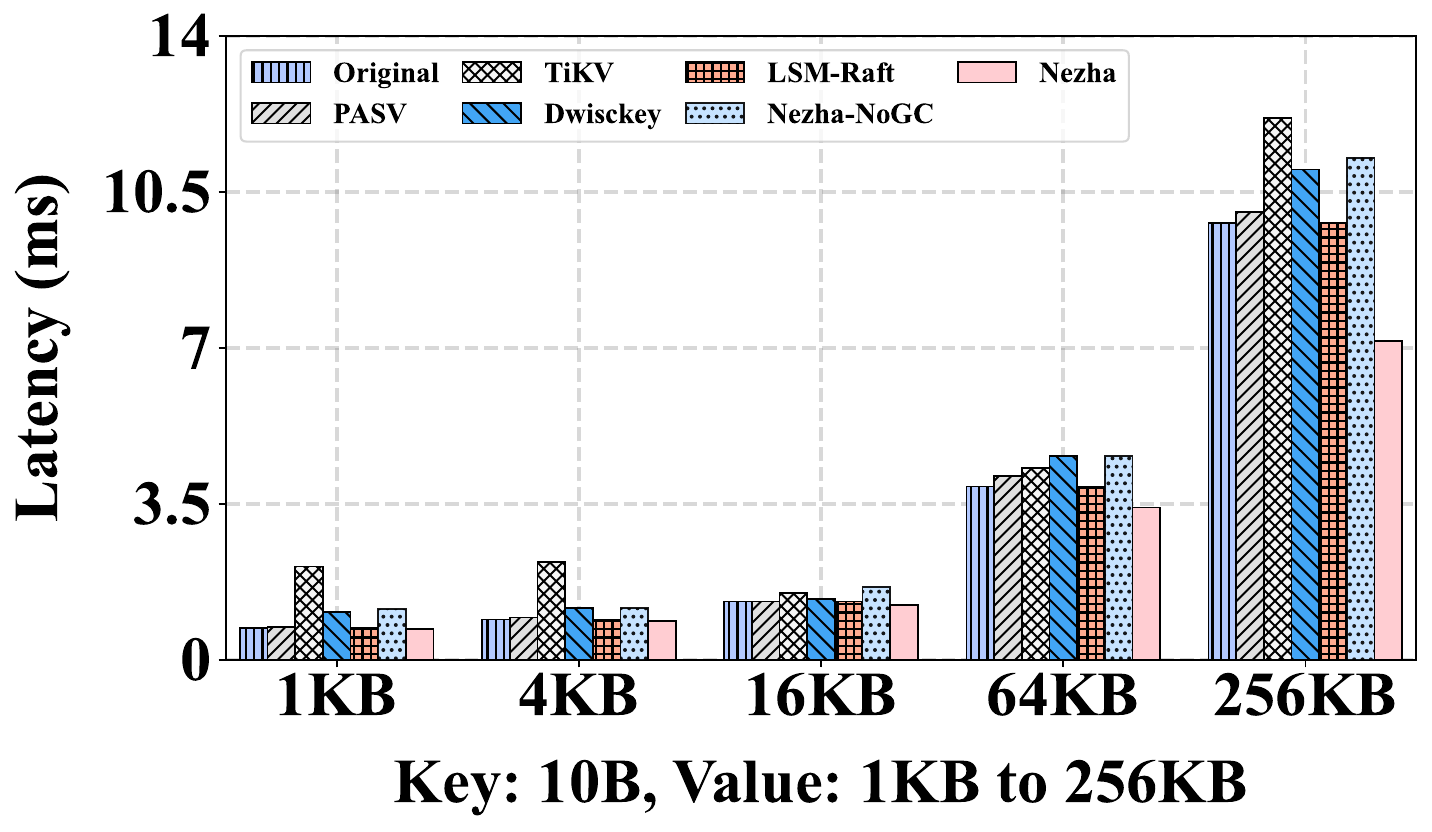}
            \label{get:latency}
        }
        %\vspace{-0.1cm}
        
        \caption{Performance comparisons for \textbf{point queries} under different value sizes.}
        \label{get-single}
    \end{minipage}
    \hfill
    % 第三组子图（Range查询）
    \begin{minipage}[t]{0.31\textwidth}
        \centering
        \subfigure[Throughput of \textbf{\textit{Scan}}]{
            \includegraphics[width=\textwidth]{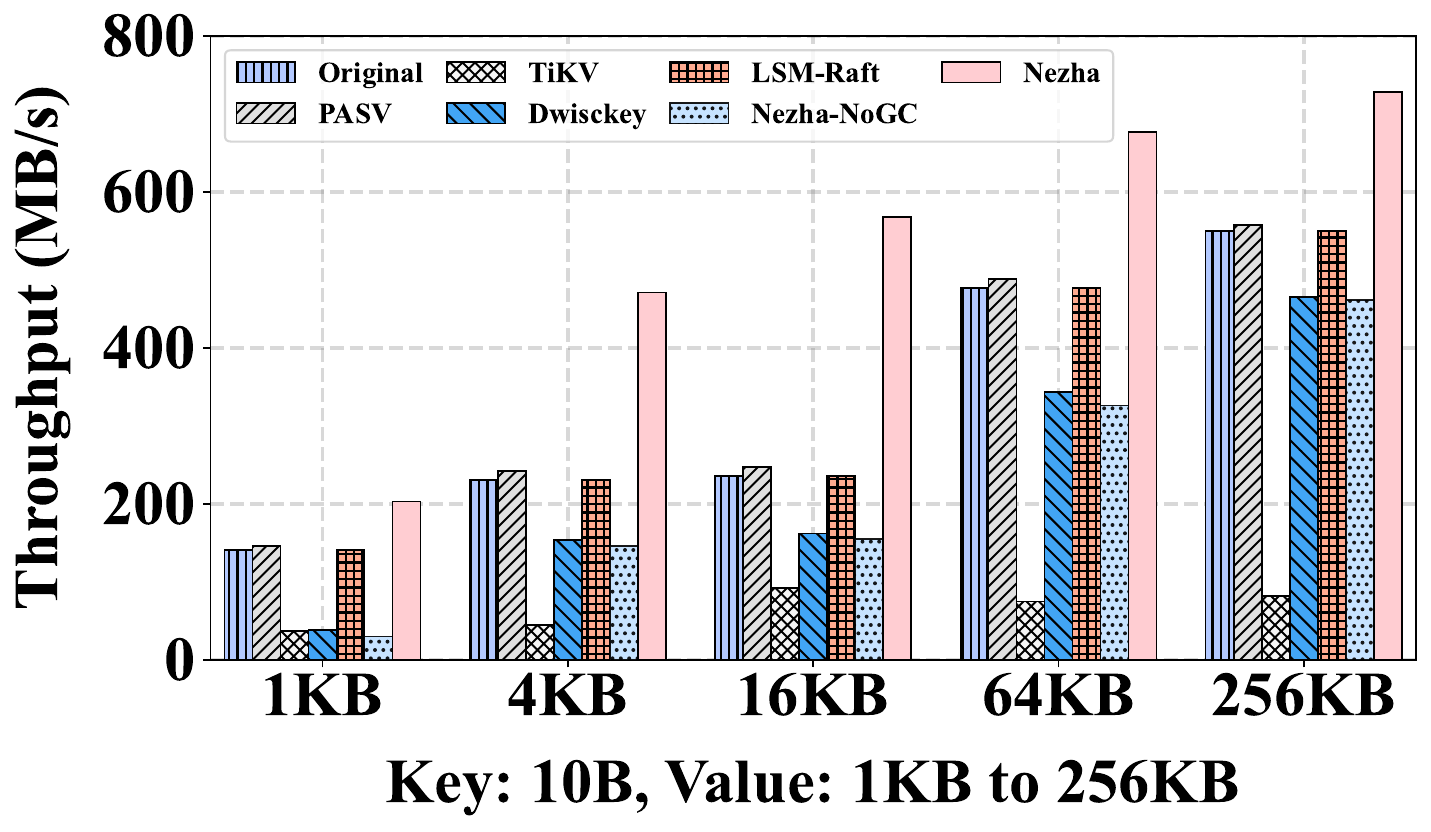}
            \label{scan:throughput}
        }
        %\vspace{0.5em}
        \vspace{-0.1cm}
        \subfigure[Average read latency of \textbf{\textit{Scan}}]{
            \includegraphics[width=\textwidth]{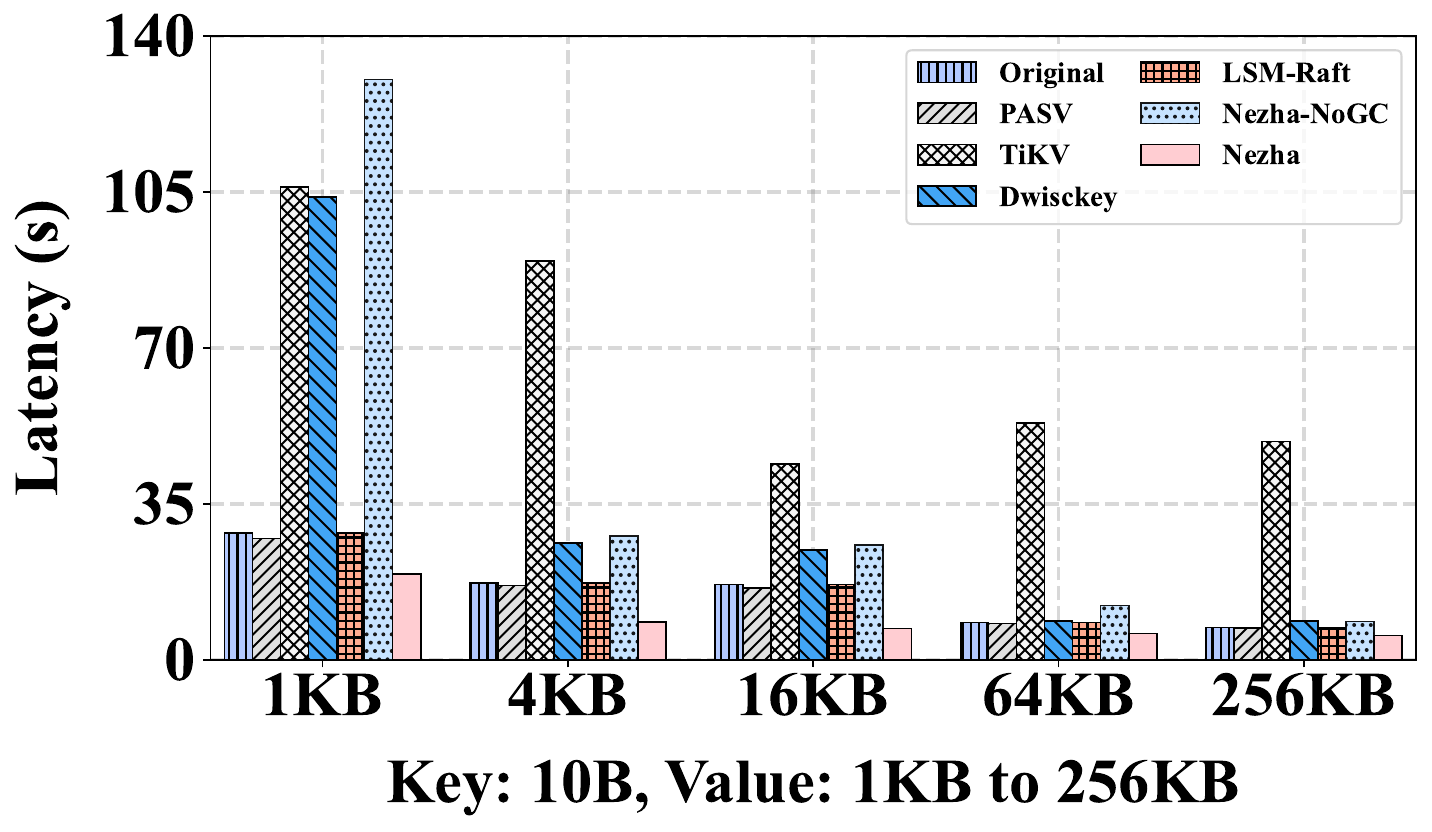}
            \label{scan:latency}
        }
        %\vspace{-0.1cm}
        
        \caption{Performance comparisons for \textbf{range queries} under different value sizes; 4GB of data is queried from a 100 GB dataset.}
        \label{scan-single}
    \end{minipage}
    \vspace{-0.5cm}
\end{figure*}

\subsubsection{\textbf{Put Performance Comparison}}

\textbf{Throughput}.
As shown in Figure~\ref{put-single} (a), the throughput of all systems demonstrates an upward trend as the value size increases from 1 KB to 256 KB. Notably, the performance advantage of Nezha and Nezha-NoGC becomes increasingly pronounced with larger value sizes. This trend is expected because larger values amplify the write amplification overhead in traditional systems, where each value must be persisted at least three times (Raft log, WAL, and SSTable). In contrast, Nezha's key-value separation architecture persists values exactly once in the ValueLog, effectively eliminating this redundant I/O overhead.

On average, Nezha achieves a 460.2\% throughput improvement over Original, with Nezha-NoGC showing similar gains (464.7\%). The minimal difference between Nezha and Nezha-NoGC indicates that background GC has negligible impact on write performance.

Among other baselines, PASV achieves modest improvement (26.5\%) over Original by eliminating the storage engine's WAL, but still suffers from Raft log and SSTable redundancy. Dwisckey, as a distributed key-value separation system, shows performance close to Nezha-NoGC but slightly lower (7.5\%) due to its additional value persistence operation. TiKV, an enterprise-level system, follows the traditional Raft-LSM architecture and thus exhibits similar limitations as Original.

LSM-Raft achieves a 16.9\% improvement over Original by reducing follower-side compaction overhead. However, Nezha outperforms LSM-Raft by 418\% because the leader, not the followers, dominates the critical write path. Nezha addresses this bottleneck by reducing redundant writes on both leader and followers—from at least three times to exactly once.

\textbf{Latency}.
As shown in Figure~\ref{put-single} (b), the latency trends are consistent with the throughput observations. On average, Nezha achieves a 59.2\% latency reduction compared to Original, while Nezha-NoGC shows a similar reduction of 60.76\%. Notably, even though Nezha performs two GC operations during the 100 GB data loading process, its write latency remains nearly identical to Nezha-NoGC, indicating that background GC has minimal impact on write performance. A detailed analysis of GC's impact is provided in Section~\ref{section4.6}.

\subsubsection{\textbf{Get Performance Comparison}}
\textbf{Throughput}.
The get operation performance is evaluated by issuing 1,000,000 point queries against the loaded 100 GB dataset. As shown in Figure~\ref{get-single} (a), key-value separation exhibits a dual effect on read performance. Nezha-NoGC shows a 21.3\% lower throughput compared to Original on average, because each point query requires an additional offset lookup before retrieving the actual value. This overhead is particularly pronounced with smaller values, as the offset lookup occupies a larger portion of the total query latency.

In contrast, Nezha achieves a 12.5\% throughput improvement over Original by leveraging its GC mechanism, which reorganizes data into sorted files with hash-based indexing. The hash index enables direct offset lookups, effectively compensating for the inherent overhead of key-value separation. This advantage becomes more significant with larger values, where the optimized data layout provides greater benefits.

Among other baselines, PASV and LSM-Raft exhibit read performance comparable to Original, as their optimizations focus on the write path and do not modify read operations. Dwisckey shows similar performance to Nezha-NoGC since both systems use similar mechanisms
for key-value separation and data retrieval. On average, Nezha outperforms Dwisckey by 37.34\%, demonstrating the effectiveness of Nezha's GC mechanism in addressing the read performance challenges inherent in key-value separation architectures.

% \textbf{Latency}.
% As shown in Figure~\ref{get-single} (b), the point query latency of all systems demonstrates an upward trend as the value size increases from 1 KB to 256 KB. On average, latency measurements show Nezha-NoGC with 28.7\% higher response times compared to the Original system, while Nezha achieves a 10.3\% reduction. This reduction in Nezha's latency can be attributed to its implementation of hash indexing that optimizes data access patterns.

% PASV demonstrates read latency comparable to that of Original across all value sizes. On average, Nezha reduces latency by 12.95\% compared to PASV. Nezha shows better read performance compared to TiKV, with an average latency reduction of 41.24\%. 

% Dwisckey's read performance is similar to Nezha-NoGC across most value sizes. On average, Nezha achieves a 25.5\% latency reduction compared to Dwisckey.

% {\color{R4} \textbf{[R4-O2]} 
% LSM-Raft exhibits read latency that is largely comparable to Original, indicating that its write-path optimizations have limited impact on read performance. In contrast, Nezha consistently achieves lower read latency than LSM-Raft, with the advantage becoming more pronounced as value sizes increase. This improvement is attributed to Nezha’s hash-based indexing, which enables direct offset lookups and avoids the multi-level search overhead inherent in traditional LSM-tree architectures.
% }
\textbf{Latency}.
As shown in Figure~\ref{get-single} (b), the latency trends are consistent with the throughput observations. On average, Nezha-NoGC exhibits 28.7\% higher latency compared to Original due to the additional offset lookup overhead, while Nezha achieves a 10.3\% latency reduction through its hash-based indexing that enables direct offset lookups.

%\vspace{-0.6cm}

\subsubsection{\textbf{Scan Performance Comparison}}
\textbf{Throughput}.
As shown in Figure~\ref{scan-single} (a), key-value separation has a more pronounced impact on range queries compared to point queries. Nezha-NoGC shows a 39.5\% lower throughput compared to Original on average, significantly worse than the 21.3\% degradation observed for point queries. This is because range queries require retrieving multiple consecutive key-value pairs, but key-value separation scatters values across the ValueLog in arrival order, converting sequential I/O into random I/O.

% In contrast, Nezha achieves a 72.6\% throughput improvement over Original, substantially higher than the 12.5\% improvement for point queries. This significant gain stems from Nezha's GC mechanism, which reorganizes scattered values into sorted files based on key order, restoring sequential access patterns for range queries. The performance gap between Nezha-NoGC and Nezha narrows with larger value sizes, as the difference between sequential and random I/O diminishes when individual I/O operations become larger.
In contrast, Nezha achieves a 72.6\% throughput improvement over Original. This significant gain stems from Nezha's GC mechanism, which reorganizes scattered values into sorted files based on key order. With the hash index, a range query only requires one random read to locate the starting position in the sorted ValueLog, followed by sequential reads for subsequent key-value pairs. This approach is more efficient than Original, which must traverse multiple SSTable files across different LSM-tree levels. The performance gap between Nezha-NoGC and Nezha narrows with larger value sizes, as the difference between sequential and random I/O diminishes when individual I/O operations become larger.

Among other baselines, PASV and LSM-Raft exhibit scan performance comparable to Original since their write-path optimizations do not improve data layout for read operations. Dwisckey shows similar performance to Nezha-NoGC as both systems lack read-optimized data reorganization. On average, Nezha outperforms Dwisckey by 208.9\%, demonstrating that Nezha's GC mechanism effectively addresses the scan performance degradation typically associated with key-value separation.

% \textbf{Latency}.
% As shown in Figure~\ref{scan-single} (b), the range query latency of all systems demonstrates a downward trend as the value size increases from 1 KB to 256 KB. For range query operations, our measurements show: Nezha-NoGC increases response times by 107.2\% compared to the Original system, while Nezha reduces scan latency by a significant 39.2\%. This reduction in range query latency demonstrates that Nezha effectively addresses the scan operation slowdown typically associated with key-value separation architectures.

% PASV and Original exhibit very similar range query latencies across all value sizes. Nezha demonstrates significantly better scan performance compared to PASV, with an average latency reduction of 37.1\%. Compared to TiKV, Nezha shows an average latency reduction of 86.8\%.

% Dwisckey demonstrates range query latencies similar to Nezha-NoGC across most value sizes. On average, Nezha achieves a 58.1\% latency reduction compared to Dwisckey.

% {\color{R4} \textbf{[R4-O2]} 
% LSM-Raft exhibits latency behavior largely comparable to Original, indicating that write-path optimizations alone provide limited benefits for read-intensive workloads. In contrast, Nezha consistently achieves lower latency than LSM-Raft by addressing the fundamental inefficiencies of LSM-tree range queries through its sorted file organization, which significantly improves access locality.
% }
\textbf{Latency}.
As shown in Figure~\ref{scan-single} (b), the latency trends are consistent with the throughput observations. On average, Nezha-NoGC increases latency by 107.2\% compared to Original due to the random I/O overhead from scattered values, while Nezha achieves a 39.2\% latency reduction through its sorted file organization that restores sequential access patterns. These results confirm that Nezha effectively addresses the scan performance degradation inherent in key-value separation architectures.

\vspace{-0.2em}
\subsection{Performance Analysis Based on Scan Length}
\label{section4.4}
To evaluate the impact of scan cardinality on range query performance, we conducted a comprehensive analysis measuring scan throughput and latency across varying query sizes. Our experiments systematically tested range queries retrieving 10, 100, 1000, and 10000 key-value pairs to understand scalability characteristics of each system. 
For these tests, we chose the middle value size of 16 KB and used 100 concurrent threads to evaluate systems, consistent with the YCSB workload \textit{E} configuration in Section~\ref{section4.5}.

\begin{figure}[tp]
\centerline{\includegraphics[width=0.48\textwidth]{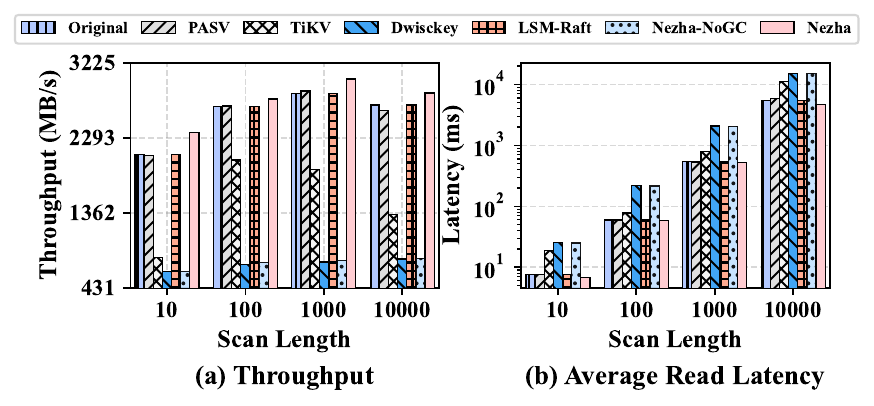}}
    \vspace{-0.4cm}
    \caption{Performance comparisons for \textbf{range queries} under different scan length.}
    \label{scan-length-16KB}
    \vspace{-0.5cm}
\end{figure}

\textbf{Throughput}. 
As shown in Figure~\ref{scan-length-16KB} (a), this experiment evaluates the impact of scan cardinality on range query performance by varying the number of retrieved key-value pairs from 10 to 10000. The results demonstrate that Nezha consistently outperforms Original across all scan lengths, achieving an average throughput improvement of 7.58\%. This stable performance advantage confirms that Nezha's GC mechanism provides reliable optimization for range queries regardless of scan size. The performance trends of other baselines are consistent with observations from Section~\ref{section4.3}, where PASV and LSM-Raft show similar performance to Original, while Dwisckey exhibits similar behavior to Nezha-NoGC.

%\subsubsection{Latency}
\textbf{Latency}.
As shown in Figure~\ref{scan-length-16KB} (b), the latency trends are consistent with the throughput observations. Nezha maintains stable and low latency across all scan lengths, while Nezha-NoGC exhibits significantly higher latency due to the overhead of key-value separation without GC optimization. These results further confirm that Nezha's GC mechanism effectively addresses the read performance degradation inherent in key-value separation architectures.

% 第四组子图（不同长度的Range查询）
% \begin{figure}[!t]
%     \centering
%     \begin{minipage}[t]{0.4\textwidth}
%         \centering
%         \includegraphics[width=\textwidth]{Figures/FIG7-a.pdf}
%         \caption*{(a) Throughput}
%         \label{scan:throughput-16KB}
%     \end{minipage}
%     \hfill
%     \begin{minipage}[t]{0.4\textwidth}
%         \centering
%         \includegraphics[width=\textwidth]{Figures/FIG7-b.pdf}
%         \caption*{(b) Average read latency}
%         \label{scan:latency-16KB}
%     \end{minipage}
%     \vspace{-0.2cm}
%     \caption{Performance comparisons for \textbf{range queries} under different scan length.}
%     \label{scan-length-16KB}
%     \vspace{-0.3cm}
% \end{figure}

% \begin{figure}[!t]
%     \centering
%     \subfigure[Throughput]{
%     \includegraphics[width=0.45\columnwidth]{Figures/FIG7-a.pdf}
%     \label{scan:throughput-16KB}
%     }
%     % \hfill
%     \subfigure[Average read latency]{
%     \includegraphics[width=0.45\columnwidth]{Figures/FIG7-b.pdf}
%     \label{scan:latency-16KB}
%     }
%     \caption{Performance comparisons for \textbf{range queries} under different scan length.}
%     \label{scan-length-16KB}
%     \vspace{-0.3cm}
% \end{figure}

% \begin{figure}[!t]
%     \centering
%     \begin{minipage}[t]{0.42\textwidth}
%         \centering
%         \includegraphics[width=\textwidth]{Figures/FIG10-a.pdf}
%         \caption*{(a) Throughput}
%         \label{GC:throughput-16KB}
%     \end{minipage}
%     \hfill
%     \begin{minipage}[t]{0.42\textwidth}
%         \centering
%         \includegraphics[width=\textwidth]{Figures/FIG10-b.pdf}
%         \caption*{(b) Average read latency}
%         \label{GC:latency-16KB}
%     \end{minipage}
%     \caption{Performance comparison of throughput and latency among original, Nezha, and Nezha-NoGC.}
%     \label{GC-16KB}
%     %\vspace{-0.5cm}
% \end{figure}

% \begin{figure}[tbp]
%     \centerline{\includegraphics[width=0.42\textwidth]{Figures/FIG9.pdf}}
%     \caption{Performance comparisons for recovery under different system states.}
%     \label{FIG9}
% \end{figure}
% 第五组子图（YCSB测试）
\begin{figure*}[!t]
    \centering
    \subfigure[Throughput]{
    \includegraphics[width=0.31\textwidth]{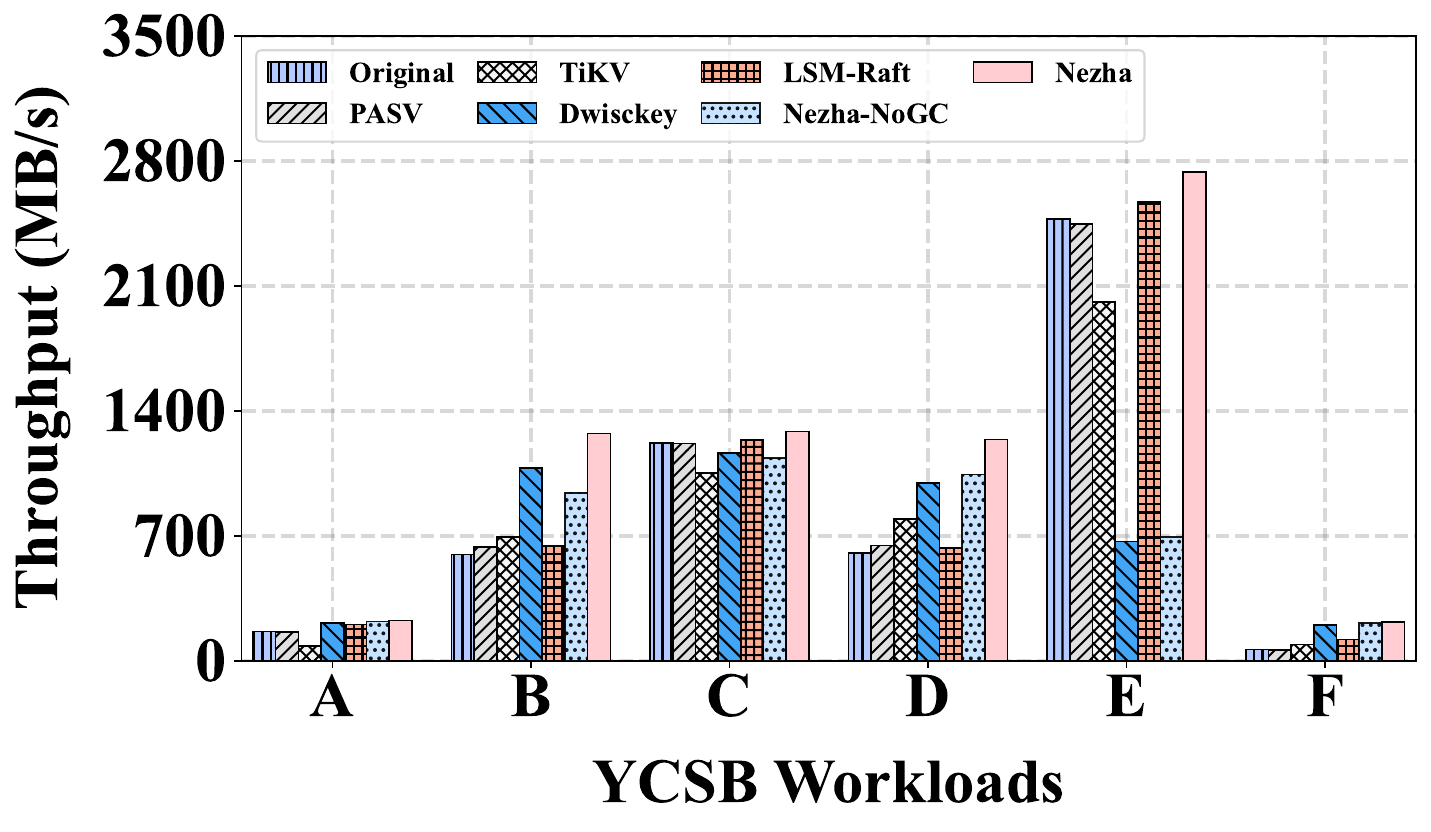}
    \label{ycsb:throughput-16KB}
    }
    \hfill
    \subfigure[Average write latency]{
    \includegraphics[width=0.31\textwidth]{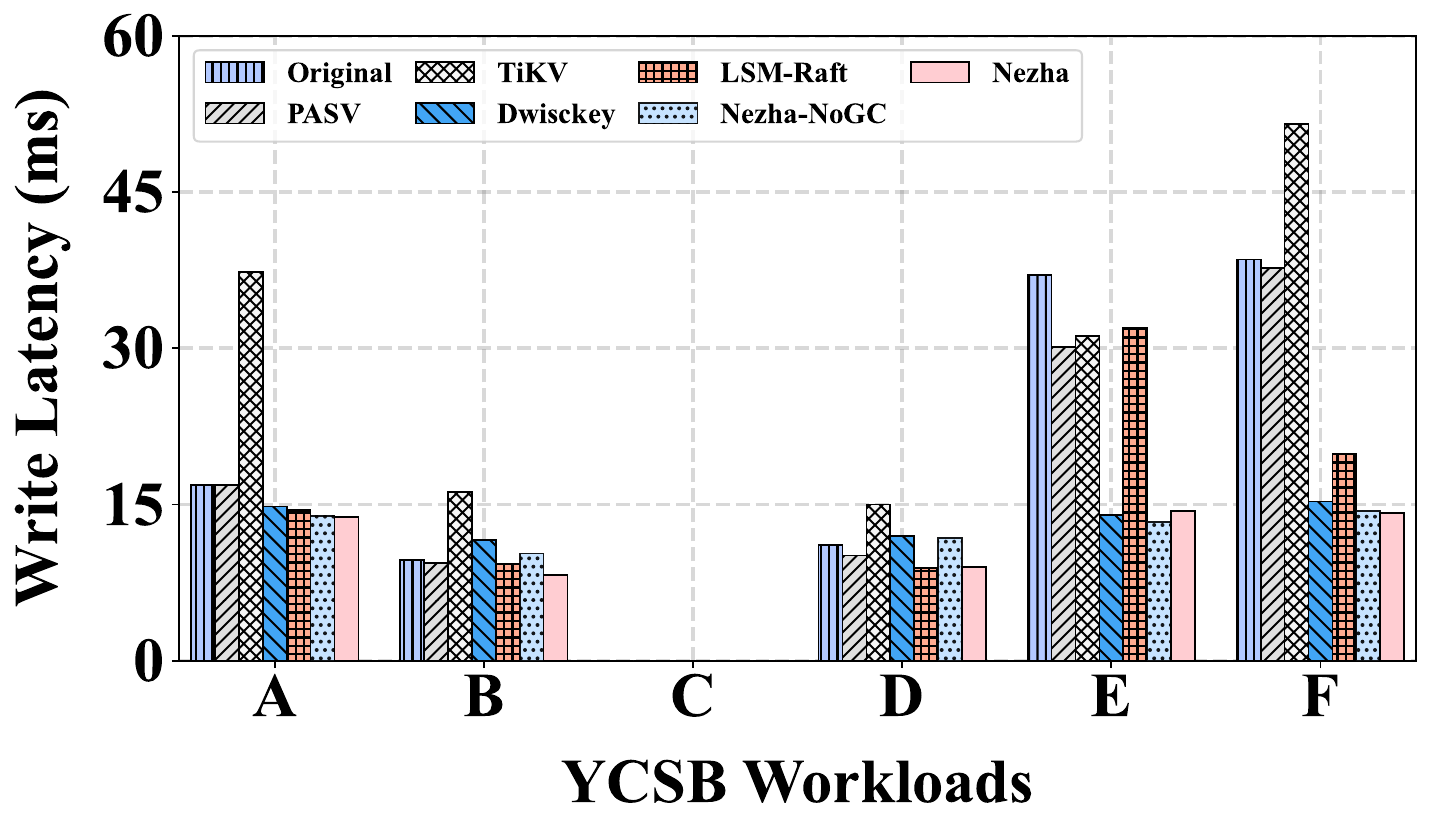}
    \label{ycsb:writelatency-16KB}
    }
    \subfigure[Average read latency]{
    \includegraphics[width=0.31\textwidth]{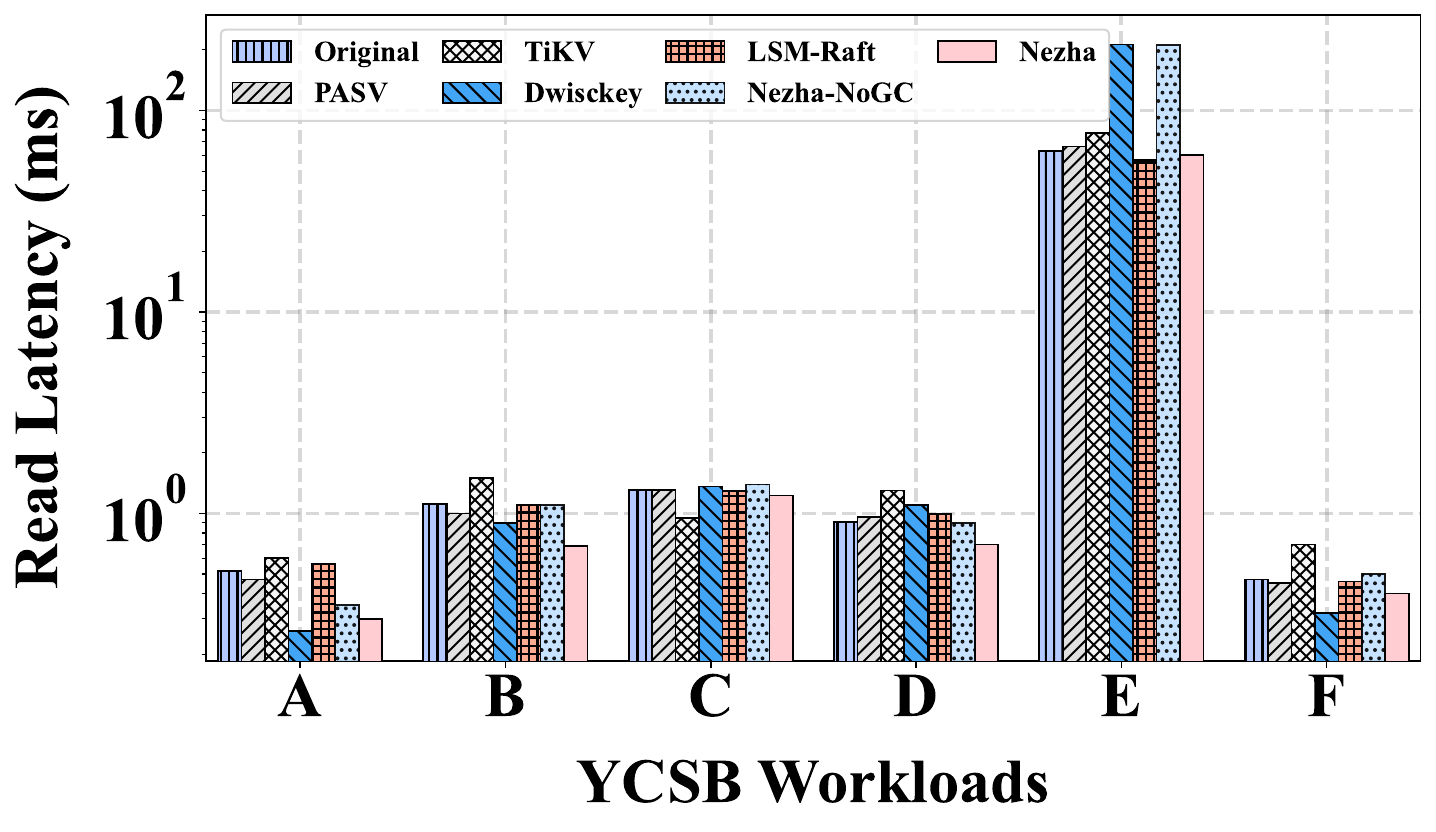}
    \label{ycsb:readlatency-16KB}
    }
    \vspace{-0.1cm}
    \caption{Throughput and latency comparisons among all systems under \textbf{YCSB} workloads.}
    \vspace{-0.3cm}
    \label{ycsb-16KB}
\end{figure*}

%\vspace{-0.1cm}
\subsection{YCSB Workload Performance}
\label{section4.5}
To validate system stability across different phases under varied read-write ratio workloads and compare performance with all baseline systems, we conducted comprehensive tests using YCSB's six classic workloads (\textit{A-F}). We pre-loaded 100 GB of random data. 
%and configured the GC mechanism to trigger when the active storage module reaches 40 GB, which represents a more realistic application scenario. 
Subsequently, we executed 1,000,000 requests for each workload type. For consistency with our previous experiments, we used a value size of 16 KB for all workloads.

\textbf{Throughput}. 
As shown in Figure~\ref{ycsb-16KB} (a), this experiment evaluates system performance under YCSB's classic workloads that simulate real-world scenarios with varying read-write ratios. The results demonstrate that Nezha consistently outperforms Original across all workload patterns, achieving an average throughput improvement of 86.5\%.

Analyzing by workload type, for write-intensive workloads (\textit{A} and \textit{F} with 50\% writes), both Nezha and Nezha-NoGC show substantial improvements due to the benefits of key-value separation. For read-dominant workloads (\textit{B}, \textit{C}, and \textit{D} with 95-100\% point queries), Nezha maintains strong performance through its hash-indexed sorted files, while Nezha-NoGC shows limited gains due to the offset lookup overhead. For scan-heavy workload \textit{E}, Nezha achieves significant improvement whereas Nezha-NoGC suffers severe degradation, consistent with observations from Section~\ref{section4.3}.

These results confirm that Nezha effectively balances read and write performance across diverse workload patterns, making it suitable for real-world applications with mixed operations. The performance trends of other baselines are consistent with previous experiments, where PASV and LSM-Raft show similar performance to Original, while Dwisckey exhibits similar behavior to Nezha-NoGC.

%\subsubsection{Write Latency} 
\textbf{Latency}. 
Figure~\ref{ycsb-16KB} (b) and Figure~\ref{ycsb-16KB} (c) present the average write latency and read latency across different YCSB workloads, respectively. The results demonstrate that Nezha consistently achieves lower latency compared to baseline systems across all workloads. For write operations, Nezha shows substantial latency reductions, particularly in write-intensive workloads. For read operations, Nezha maintains competitive latency performance across diverse workload patterns, including the challenging scan-heavy workload \textit{E} where Nezha-NoGC suffers significant degradation. These observations confirm that Nezha effectively balances both write and read performance under realistic mixed workloads.

\subsection{Scalability Analysis}
\label{scalability}
To address the scalability characteristics of Nezha under larger cluster configurations, we conducted experiments with 3, 5, and 7 node clusters. These experiments evaluate system scalability by writing 100 GB of data using a fixed value size of 16 KB, consistent with our previous experiments.

% \begin{figure}[!t]
%     \centering
%     \vspace{-0.4cm}
%     \subfigure[Throughput]{%
%     \includegraphics[width=0.48\columnwidth]{Figures/FIG_scalability_throughput.pdf}%
%     \label{scalability:throughput-16KB}%
%     }%
%     \subfigure[Average write latency]{%
%     \includegraphics[width=0.48\columnwidth]{Figures/FIG_scalability_latency.pdf}%
%     \label{scalability:latency-16KB}%
%     }
%     \caption{Scalability analysis: Put throughput and latency across different cluster sizes (3, 5, and 7 nodes) with 16 KB values.}
%     \label{scalability-16KB}
%     \vspace{-0.4cm}
% \end{figure}

\begin{figure}[tp]
\centerline{\includegraphics[width=0.48\textwidth]{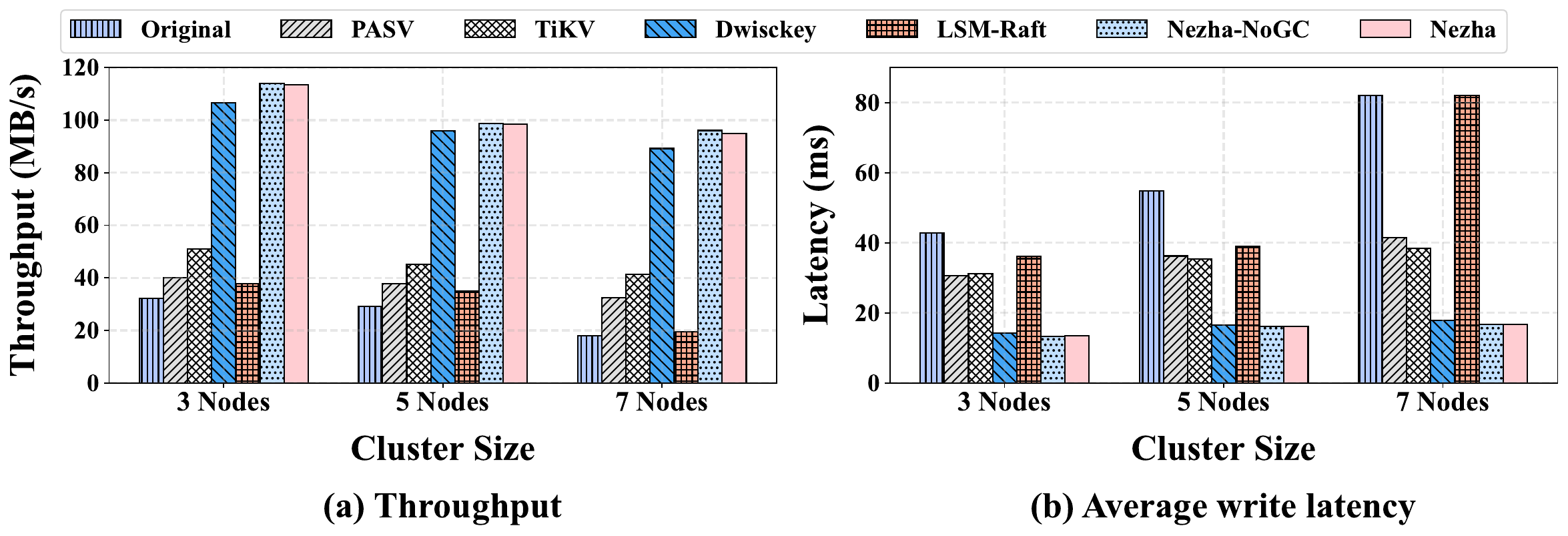}}
    \vspace{-0.2cm}
    \caption{Performance comparisons for \textbf{put} requests under different cluster sizes.}
    \label{scalability-16KB}
    \vspace{-0.4cm}
\end{figure}

\textbf{Throughput and Latency}.
As shown in Figure~\ref{scalability-16KB}, all systems exhibit decreased throughput and increased latency as cluster size grows from 3 to 7 nodes due to consensus coordination overhead. Nezha demonstrates substantially better scalability, achieving 3.5$\times$ to 5.3$\times$ higher throughput and 3.2$\times$ to 4.9$\times$ lower latency than Original across different cluster sizes. These results confirm that Nezha maintains robust performance regardless of cluster size.

% Notably, Nezha's performance advantage over Original amplifies as cluster size increases: the throughput ratio improves from 3.5$\times$ at 3 nodes to 3.4$\times$ at 5 nodes and 5.3$\times$ at 7 nodes. This widening gap demonstrates that Nezha's architectural benefits become more pronounced under more demanding distributed conditions.

% \textbf{Latency}.
% As shown in Figure~\ref{fig:scalability} (b), latency trends mirror the throughput observations. Original's write latency increases dramatically from 42.8 ms (3 nodes) to 82.05 ms (7 nodes), a 91.7\% increase. LSM-Raft exhibits similar behavior, with latency rising from 36.2 ms to 82.01 ms (126.5\% increase).

% Nezha maintains significantly more stable latency characteristics across cluster sizes. Its write latency increases from 13.4 ms (3 nodes) to 16.8 ms (7 nodes), representing only a 25.4\% increase. At 7 nodes, Nezha achieves 4.9$\times$ lower latency than Original and 4.9$\times$ lower than LSM-Raft.

% These results demonstrate that Nezha's key-value separation strategy not only improves absolute performance but also enhances system scalability. The architectural decision to persist values only once in the ValueLog, rather than propagating complete key-value pairs through multiple persistence layers, effectively mitigates the consensus overhead that traditionally limits the scalability of Raft-based storage systems.

\subsection{Impact of GC on Performance}
\label{section4.6}
To quantify the specific impact of the GC mechanism on system performance, we designed a targeted long-duration write test. In this experiment, we set the GC trigger threshold to 40 GB and continuously monitored the performance of three systems during the process of writing 100 GB of data: Original, Nezha, and Nezha-NoGC. For consistency with our previous experiments, we used a value size of 16 KB for all baselines, with performance snapshots collected every 100 ms.

\begin{figure}[!t]
    \centering
    \vspace{-0.5cm}
    \subfigure[Throughput]{%
    \includegraphics[width=0.48\columnwidth]{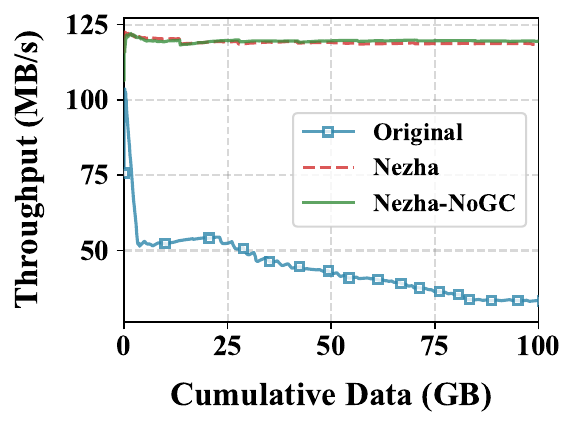}%
    \label{GC:throughput-16KB}%
    }%
    \subfigure[Average write latency]{%
    \includegraphics[width=0.48\columnwidth]{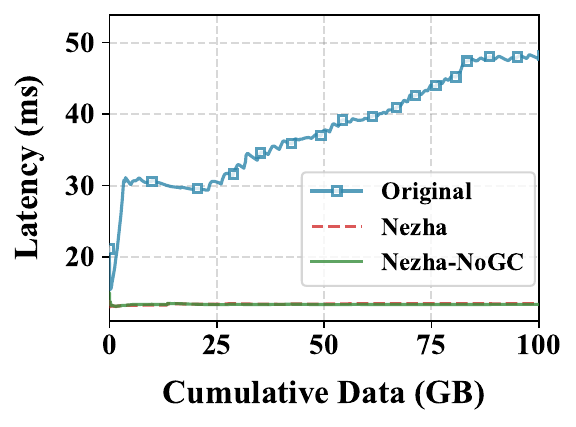}%
    \label{GC:latency-16KB}%
    }
    \caption{Performance impact of GC on the system.}
    \label{GC-16KB}
    \vspace{-0.5cm}
\end{figure}

\textbf{Throughput}.
% As shown in Figure~\ref{GC-16KB} (a), two GC operations are triggered during the 100 GB data write process (at 40 GB and 80 GB data points, respectively). From the cumulative throughput curves, we observe that despite enabling the GC mechanism, the performance gap between Nezha and Nezha-NoGC remains minimal throughout the test, with their throughput curves nearly overlapping. This demonstrates that our Raft-aware GC mechanism controls the performance overhead within an acceptable range. In contrast, the Original system shows significantly inferior performance, validating the superiority of the key-value separation architecture in handling large-scale data writes.
As shown in Figure~\ref{GC-16KB} (a), two GC operations are triggered during the 100 GB data write process (at 40 GB and 80 GB data points, respectively). From the cumulative throughput curves, we observe that despite enabling the GC mechanism, the performance gap between Nezha and Nezha-NoGC remains minimal throughout the test, with their throughput curves nearly overlapping.
This is because Nezha atomically switches write requests to the New Storage module when GC is triggered, while GC operations execute asynchronously on the separate Active Storage module, effectively decoupling GC overhead from the critical write path. 
In contrast, the Original system shows significantly inferior performance, validating the superiority of the key-value separation architecture in handling large-scale data writes.

\textbf{Latency}.
Figure~\ref{GC-16KB} (b) presents the latency evolution trends during the data writing process. The latency analysis confirms similar findings: Nezha and Nezha-NoGC exhibit nearly identical latency performance concentrated in the lower range, while the Original system shows higher and more dispersed latency distribution. This difference stems from Nezha's key-value separation architecture that reduces write amplification and avoids the compaction-induced latency spikes inherent in traditional LSM-Tree designs.

% \begin{figure}[tp]
% \centerline{\includegraphics[width=0.5\textwidth]{Figures/combined_three_charts.pdf}}
%     \vspace{-0.2cm}
%     \caption{GC Performance Impact and Recovery Time under Different System States.}
%     \label{GC-16KB}
%     \vspace{-0.4cm}
% \end{figure}

\subsection{Fault and Recovery}
\label{section4.7}
\begin{figure}[tp]
\centerline{\includegraphics[width=0.3\textwidth]{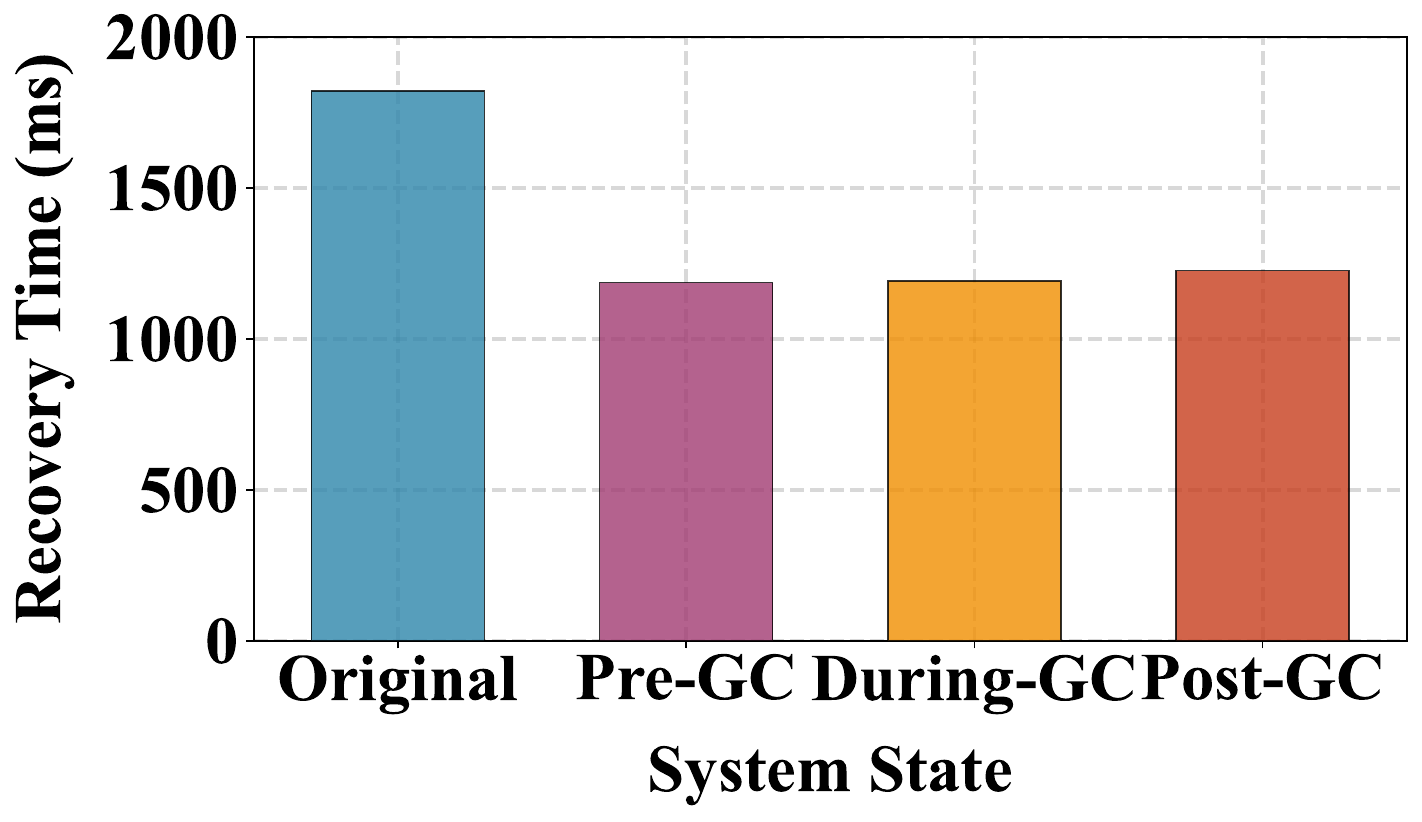}}
    \vspace{-0.4cm}
    \caption{Recovery time comparisons for different system states.}
    \label{ft}
    \vspace{-0.5cm}
\end{figure}

To quantify the impact of GC on failure recovery time, we conducted recovery experiments across different system states. Figure~\ref{ft} shows the recovery time comparison, where Pre-GC, During-GC, and Post-GC phases achieved 34.8\%, 34.5\%, and 32.6\% reductions in recovery time compared to the Original system, respectively. This improvement is attributed to the introduction of key-value separation, which significantly reduces the data volume stored in RocksDB by storing only lightweight offset pointers instead of complete values. Even when failures occur during GC, the recovery process only requires an additional step of reading the interrupt point from the sorted ValueLog to complete the remaining GC process.
%\vspace{0.1cm}

\section{RELATED WORK}
\label{sectio5}
 
% Since Nezha addresses the performance bottleneck arising from the interaction between consensus protocols and storage engines, we organize related work into three categories: (A) recent advances in log-structured KV stores, (B) techniques for write amplification reduction, and (C) consensus-storage co-design and optimization. We discuss recent advances in each category and highlight how Nezha differs from existing approaches.
Motivated by performance bottlenecks at the consensus–storage interface, we review recent work on log-structured KV stores, write-amplification reduction, and consensus–storage co-design, and discuss how Nezha differs from existing approaches.

\subsection{Recent Advances in Log-Structured KV Stores}
LSM-tree-based key-value stores have become the foundation of modern storage systems due to their write-optimized design \cite{b20,b21,b22}. Recent developments have focused on improving I/O efficiency and adapting LSM-trees to emerging hardware and architectural paradigms. On the hardware front, researchers have sought to optimize data placement and reduce maintenance overhead by adapting LSM-trees for ZNS SSDs through lifetime-aware data management \cite{b80}, exploiting GPU offloading for performance-critical operations \cite{b81}, targeting computational storage drives with adaptive host-CSD offloading \cite{b83}, and leveraging large DRAM capacities with memory-efficient indexing \cite{b82}. On the architecture front, cloud-native deployment requirements have driven adaptations for disaggregated memory \cite{b55}, serverless environments \cite{b57}, data confidentiality under multi-tenant constraints \cite{b77}, and object storage-based ``diskless'' designs that treat cloud stores as the primary persistence layer \cite{b84}. Additionally, learning-based approaches \cite{b78,b79} enable adaptive parameter tuning under dynamic workloads.

However, these works primarily optimize LSM-tree internals to accommodate evolving system characteristics and deployment environments, without systematically restructuring the interaction with consensus protocols. In contrast, Nezha addresses cross-layer redundancy through consensus-storage co-design, eliminating redundant persistence by integrating key-value separation directly into the Raft protocol. Since Nezha operates at the consensus layer without modifying LSM-tree internals, these hardware and architecture adaptations remain \textit{orthogonal} to our approach and could be combined with Nezha for further performance gains.

% \vspace{-0.7em}
\subsection{Techniques for Write-Amplification Reduction}
Beyond hardware and architecture adaptations, write amplification remains a fundamental performance challenge in LSM-tree-based storage systems. This problem has been addressed from two complementary directions. Key-value separation, pioneered by WiscKey \cite{b12}, reduces write amplification by storing only keys in the LSM-tree while placing values in a separate append-only log, thereby avoiding the repeated movement of large values during compaction. Compaction, as another dominant source of write amplification, has received extensive attention through workload-aware tuning \cite{b52}, hardware offloading \cite{b53,b63,b68}, pipelining parallelism \cite{b64}, resource-aware scheduling \cite{b56,b65}, and file selection refinement \cite{b75}.

However, these works optimize write amplification solely within single-node storage engines. Nezha addresses write amplification at a fundamentally different level by integrating key-value separation directly into the Raft consensus layer, thereby reducing redundant persistence operations across the entire distributed cluster—from at least three times to just once. Since Nezha operates at a higher abstraction level without modifying LSM-tree internals, these single-node optimizations remain orthogonal to our approach.

\vspace{-0.3em}
\subsection{Consensus-Storage Co-design and Optimization}
The co-design of consensus protocols and storage systems has gained significant attention as researchers recognize the performance overhead of redundant persistence operations. Production systems such as TiKV \cite{b48}, CockroachDB \cite{b36}, and PolarDB \cite{b38} adopt traditional layered architectures where Raft-based consensus and storage engines operate independently, inheriting redundant persistence across multiple layers. Recent research addresses this issue by eliminating redundant logging across system layers \cite{b49, b50} and optimizing consensus protocol mechanisms for shorter critical paths \cite{b11, b61}. LSM-Raft \cite{b76} bridges both directions by transmitting compacted SSTables instead of fine-grained entries, reducing follower-side redundant writes.

However, LSM-Raft's optimization benefits only followers, as leaders still experience full redundant writes. Nezha goes beyond LSM-Raft by further eliminating leader-side redundancy through architectural-level integration of key-value separation with Raft. While prior works such as PASV [28] and PALF [29] reduce at most one persistence operation, Nezha reduces value persistence from at least three times to just once while preserving Raft's safety properties. In addition, Nezha can be combined with these approaches to further reduce the persistence overhead of keys.

\vspace{-0.5em}
\section{CONCLUSION}
\label{sectio6}

% This paper proposes KVS-Raft, an innovative distributed consensus protocol optimized from the perspective of consensus algorithm layer and storage layer interaction. The core innovation of KVS-Raft lies in introducing key-value separation concepts into the Raft consensus algorithm layer, significantly reducing storage layer data redundancy and disk write overhead through redesigned interaction mechanisms between the two layers. Unlike existing work, KVS-Raft takes a macro-architectural system perspective, incorporating key-value separation mechanisms at the consensus algorithm layer, allowing optimization effects to be reflected across the entire distributed cluster. We implemented a prototype storage system based on KVS-Raft and verified its effectiveness in reducing system overhead and improving performance through extensive experiments.

In this paper, we present Nezha, a novel distributed key-value storage system that achieves comprehensive performance optimization through deep integration of consensus and storage layers.
% Unlike existing work that focuses on protocol-level improvements, Nezha takes a system-wide architectural approach to performance optimization. 
At its core, Nezha incorporates KVS-Raft, an innovative consensus protocol that integrates key-value separation concepts into the Raft consensus layer, significantly reducing storage redundancy and write overhead across the entire distributed cluster. To fully realize the system's potential, we design a Raft-aware GC framework that optimizes read performance while maintaining strong consistency guarantees. These architectural innovations, complemented by a three-phase request processing mechanism, enable Nezha to achieve substantial performance improvements over traditional Raft-based storage systems, as verified through extensive experimental evaluation.

% This paper presents Nezha, a novel distributed key-value storage system that achieves comprehensive performance optimization through system-wide architectural innovations. Unlike existing work focusing on protocol-level improvements, Nezha takes a macro-architectural perspective by deeply integrating consensus and storage layers. At its core, Nezha incorporates KVS-Raft, which introduces key-value separation mechanisms at the consensus layer, enabling optimization effects across the entire distributed cluster. To maximize system performance, we design a Raft-aware GC framework that optimizes read operations while maintaining strong consistency guarantees. Extensive experimental evaluation demonstrates that these architectural innovations, supported by a three-phase request processing mechanism, enable Nezha to achieve substantial performance improvements over traditional Raft-based storage systems.

\vspace{-0.5em}
\section*{Acknowledgment}
% This work was supported by the National Key R\&D Program of China, No. 2022YFB4501703, the National Natural Science Foundation of China (KY0402022036), and the Provincial Key Research and Development Program of Jiangxi (012031379055).
% This work is supported by the National Key R\&D Program of China under Grant 2022YFB4501703, the Jiangxi Provincial Career-Early Young Scientists and Technologists Cultivation Project under Grant 20252BEJ730003, and the Jiangxi Provincial Natural Science Foundation under Grant 20252BAC200615.
This work was supported by the National Key R\&D Program of China (2022YFB4501703), the Jiangxi Provincial Career-Early Young Scientists and Technologists Cultivation Project (20252BEJ730003), and the Jiangxi Provincial Natural Science Foundation (20252BAC200615).

% \section*{AI-Generated Content Acknowledgement}
% This paper was prepared with the assistance of Claude (Anthropic) for language polishing and refinement of the English writing. The AI tool was used to improve grammar, clarity, and readability of the manuscript text. All technical content, experimental design, implementation, and scientific contributions are solely the work of the authors.

% \balance
% \IEEEtriggeratref{30} % 第一站：从第 30 篇开始变色
% \IEEEtriggercmd{
%     \color{R4} % 开启 Reviewer #4 的颜色 (BrickRed)
%     \IEEEtriggeratref{31} % 埋伏：第 31 篇（范围结束后的第一篇）恢复
%     \IEEEtriggercmd{
%         \color{black} % 恢复黑色
%         \IEEEtriggeratref{36} % 第二站：从第 54 篇再次触发
%         \IEEEtriggercmd{
%             \color{R4} % 开启 Reviewer #1 的颜色 (blue)
            \IEEEtriggeratref{54} % 埋伏：第 56 篇恢复
            % \IEEEtriggercmd{
            %     \color{black} % 恢复黑色
            %     \IEEEtriggeratref{56} % 第三站：假设第三个范围从 60 开始
            %     \IEEEtriggercmd{
            %         \color{R4} % 开启 Reviewer #2 的颜色 (ForestGreen)
            %         \IEEEtriggeratref{56} % 埋伏：第 62 篇恢复
            %         \IEEEtriggercmd{\color{black}}
            %     }
            % }
%         }
%     }
% }
% \balance
\bibliographystyle{IEEEtran}
\bibliography{IEEEabrv,myrefs}
\end{document}